\documentclass[aps,prl,twocolumn,reprint,amsmath,amssymb,superscriptaddress,floatfix,footinbib,longbibliography]{revtex4-1}

\usepackage{epsfig}
\usepackage{graphicx}
\usepackage{epstopdf}

\usepackage[T1]{fontenc}
\usepackage[applemac]{inputenc}
\usepackage{lmodern}
\usepackage[english]{babel}

\usepackage{ae}
\usepackage{units}

\usepackage{amsmath,amssymb,natbib,bm}
\usepackage{psfrag}
\usepackage{subfigure}
\usepackage{amsthm}
\usepackage{bbm}

\usepackage{fixmath}
\usepackage{booktabs}

\usepackage[americaninductors]{circuitikz}
\usepackage{tikz}
\usetikzlibrary{arrows}

\usepackage{color}
\usepackage{url}

\usepackage[colorlinks]{hyperref}
\hypersetup{%
        plainpages=true,
        breaklinks=true,
        hypertexnames=false,
        pageanchor=true,
        colorlinks=true,
        linkcolor={blue},
        citecolor={red},
        urlcolor={blue},
        anchorcolor={black}
      }
      
\usepackage{mleftright} 

\newcommand{\figref}[1]{\mbox{Fig.~\ref{#1}}}
\newcommand{\tabref}[1]{\mbox{Table~\ref{#1}}}

\renewcommand{\eqref}[1]{\mbox{Eq.~(\ref{#1})}}

\newcommand{\comm}[2]{\mleft[ #1, #2 \mright]}
\newcommand{\lind}[1]{\mathcal{D}\mleft[#1\mright]}

\newcommand{\sz}{\sigma_z}

\newcommand{\sm}{\sigma_-}
\renewcommand{\sp}{\sigma_+}

\newcommand{\abs}[1]{\mleft|#1\mright|}

\newcommand{\nn}{\nonumber}

\newcommand{\be}{\begin{equation}}
\newcommand{\ee}{\end{equation}}
\newcommand{\bea}{\begin{eqnarray}}
\newcommand{\eea}{\end{eqnarray}}

\begin{document}

\title{Decoherence-free interaction between giant atoms in waveguide QED}

\author{Anton Frisk Kockum}
\email[e-mail:]{anton.frisk.kockum@gmail.com}
\affiliation{Center for Emergent Matter Science, RIKEN, Saitama 351-0198, Japan}

\author{G\"oran Johansson}
\affiliation{Department of Microtechnology and Nanoscience (MC2), Chalmers University of Technology, SE-412 96 Gothenburg, Sweden}

\author{Franco Nori}
\affiliation{Center for Emergent Matter Science, RIKEN, Saitama 351-0198, Japan}
\affiliation{Physics Department, The University of Michigan, Ann Arbor, Michigan 48109-1040, USA}

\date{\today}

\begin{abstract}

In quantum-optics experiments with both natural and artificial atoms, the atoms are usually small enough that they can be approximated as point-like compared to the wavelength of the electromagnetic radiation they interact with. However, superconducting qubits coupled to a meandering transmission line, or to surface acoustic waves, can realize ``giant artificial atoms'' that couple to a bosonic field at several points which are wavelengths apart. Here, we study setups with multiple giant atoms coupled at multiple points to a one-dimensional (1D) waveguide. We show that the giant atoms can be protected from decohering through the waveguide, but still have exchange interactions mediated by the waveguide. Unlike in decoherence-free subspaces, here the entire multi-atom Hilbert space is protected from decoherence. This is not possible with ``small'' atoms. We further show how this decoherence-free interaction can be designed in setups with multiple atoms to implement, e.g., a 1D chain of atoms with nearest-neighbor couplings or a collection of atoms with all-to-all connectivity. This may have important applications in quantum simulation and quantum computing.

\end{abstract}

\maketitle

\paragraph*{Introduction.}

In quantum systems, there is generally a fundamental problem of trade-off between interaction and protection from decoherence~\cite{Suter2016}. For spatially separated atoms, one way to realize a protected interaction is to use a quantum bus~\cite{Zheng2000, Blais2007}. As has been demonstrated in experiments~\cite{Osnaghi2001, Majer2007, Filipp2011}, two atoms that are detuned from a resonator (the quantum bus) to which they both couple, can interact via virtual photons in the resonator. Since the photons mediating the interaction are virtual, this interaction is protected from losses in the resonator. Although this scheme can be extended to more atoms, and interactions between more than two atoms~\cite{Stassi2017}, there are limits to the connectivity between atoms and the protection from decoherence.

Another approach to protecting quantum information is decoherence-free subspaces~\cite{Lidar1998, Lidar2014}, i.e., particular subspaces (of the total Hilbert space of a quantum system) which are less sensitive to decoherence due to the form of their coupling to the dissipative environment. A well-known example is dark (sub-radiant) states~\cite{Dicke1954}, collective atomic states which do not decay, due to interference, to the environment that the atoms are all coupled to.

One platform where it has been suggested~\cite{Paulisch2016} that such decoherence-free subspaces could be used for protected quantum computation is waveguide quantum electrodynamics (QED). In waveguide QED, atoms are coupled to, and interact via, a continuum of bosonic modes in a one-dimensional (1D) waveguide. As reviewed in Refs.~\cite{Roy2017, Gu2017}, there are many experimental realizations of waveguide QED, including quantum dots and other emitters coupled to plasmons in nanowires~\cite{Akimov2007, Huck2016}, quantum dots coupled to photonic crystal waveguides~\cite{Arcari2014}, and natural atoms coupled to optical fibers~\cite{Bajcsy2009}, but the platform with best performance is arguably superconducting artificial atoms~\cite{You2005, You2011, Gu2017} coupled to transmission lines, where several experiments have been performed in the past few years~\cite{Astafiev2010, Astafiev2010a, Abdumalikov2010, Hoi2011, Abdumalikov2011, Hoi2012, VanLoo2013, Hoi2013a, Hoi2013b, Hoi2015, Liu2017, Forn-Diaz2017, Wen2017}. There is also a wealth of theoretical work studying two~\cite{Lehmberg1970a, Milonni1974, Zhou2008a, Rist2008, Chen2011, Gonzalez-Tudela2011, Xia2011, Lalumiere2013, Zheng2013a, Derouault2014, Fratini2014, Laakso2014, Shahmoon2014, Fang2015, Douglas2016, Pichler2016} or more~\cite{Lehmberg1970, Friedberg1973, Tsoi2008, Chang2012, Stannigel2012, Lalumiere2013, Ramos2014, Sathyamoorthy2014, Fang2014, Pichler2015a, Shi2015, Guimond2016a, Mirza2016, Paulisch2016, Gonzalez-Tudela2017} atoms interacting with a 1D waveguide. For a more complete overview, see Refs.~\cite{Roy2017, Gu2017}.

Since the dark states in waveguide QED are a result of interference effects, it is relevant to explore schemes for increasing such interference. One such scheme is to terminate the waveguide with a mirror. A single atom in front of a mirror, a setup which has seen both experimental~\cite{Eschner2001, Wilson2003, Dubin2007, Hoi2015} and theoretical~\cite{Meschede1990, Dorner2002, Beige2002, Dong2009, Koshino2012, Wang2012, Tufarelli2013, Fang2015, Shi2015, Pichler2016} investigation, can be protected from decay by interference between the relaxation from the atom and its mirror image.

An implicit assumption thus far has been that the atoms are \textit{small} compared to the wavelength of the bosonic modes of the waveguide they interact with. However, interference effects can be further increased if the atoms can be \textit{giant}, a term we take to mean that the atoms can couple to the waveguide at several points, which can be spaced wavelengths apart. The physics of a single such giant atom has been explored recently~\cite{Kockum2014, Guo2017} with results including a frequency-dependent relaxation rate and Lamb shift. These works were inspired by recent experiments~\cite{Gustafsson2014, Aref2016, Manenti2017, Noguchi2017, Bolgar2017, Moores2017} realizing giant atoms by coupling superconducting artificial atoms to surface acoustic waves (SAWs), which have much shorter wavelengths than the microwaves normally used in experiments with such artificial atoms. However, as outlined in Ref.~\cite{Kockum2014}, superconducting transmission lines could be used to achieve the same effect if they are suitably meandered.

In this Letter, we present the first study of multiple giant atoms coupled to a 1D waveguide. We begin by considering the case of two giant atoms, coupled at two points each to an open waveguide, and compare this setup to two small atoms in open and semi-infinite waveguides. We show that, for a certain arrangement of the connection points of the giant atoms, decoherence into the waveguide can be \textit{completely suppressed} while the giant atoms \textit{still interact} with each other via the waveguide. Unlike the dark states for small atoms, this decoherence-free interaction is independent of the states of the giant atoms, i.e., the \textit{entire} multi-atom Hilbert space is protected from decoherence, not just a subspace.

We then generalize these results to an arbitrary number of giant atoms with an arbitrary number of connection points each. In this way, we show that protected pairwise exchange interactions between multiple giant atoms can be \textit{designed} for high connectivity (beyond nearest neighbor) and with \textit{arbitrary sign} of the coupling strengths. We outline how these setups can be implemented with superconducting circuits. 

We believe that these results can find many applications, e.g., in quantum simulation~\cite{Buluta2009, Georgescu2014}, where there is much interest in spins connected in one- or two-dimensional arrangements~\cite{Wang2006, Islam2013, Salathe2015, Giamarchi2016, Douglas2016, Bernien2017, Cedergren2017, Yang2017}. It may also be possible to use setups with giant atoms to generate entangled states such as cluster~\cite{Briegel2001} or graph~\cite{Hein2004} states, which can be used for one-way quantum computing~\cite{Raussendorf2001, Walther2005, Briegel2009}.


\paragraph*{Master equation for two atoms in a waveguide.}

\begin{figure}[]
\centering
\includegraphics[width=\columnwidth]{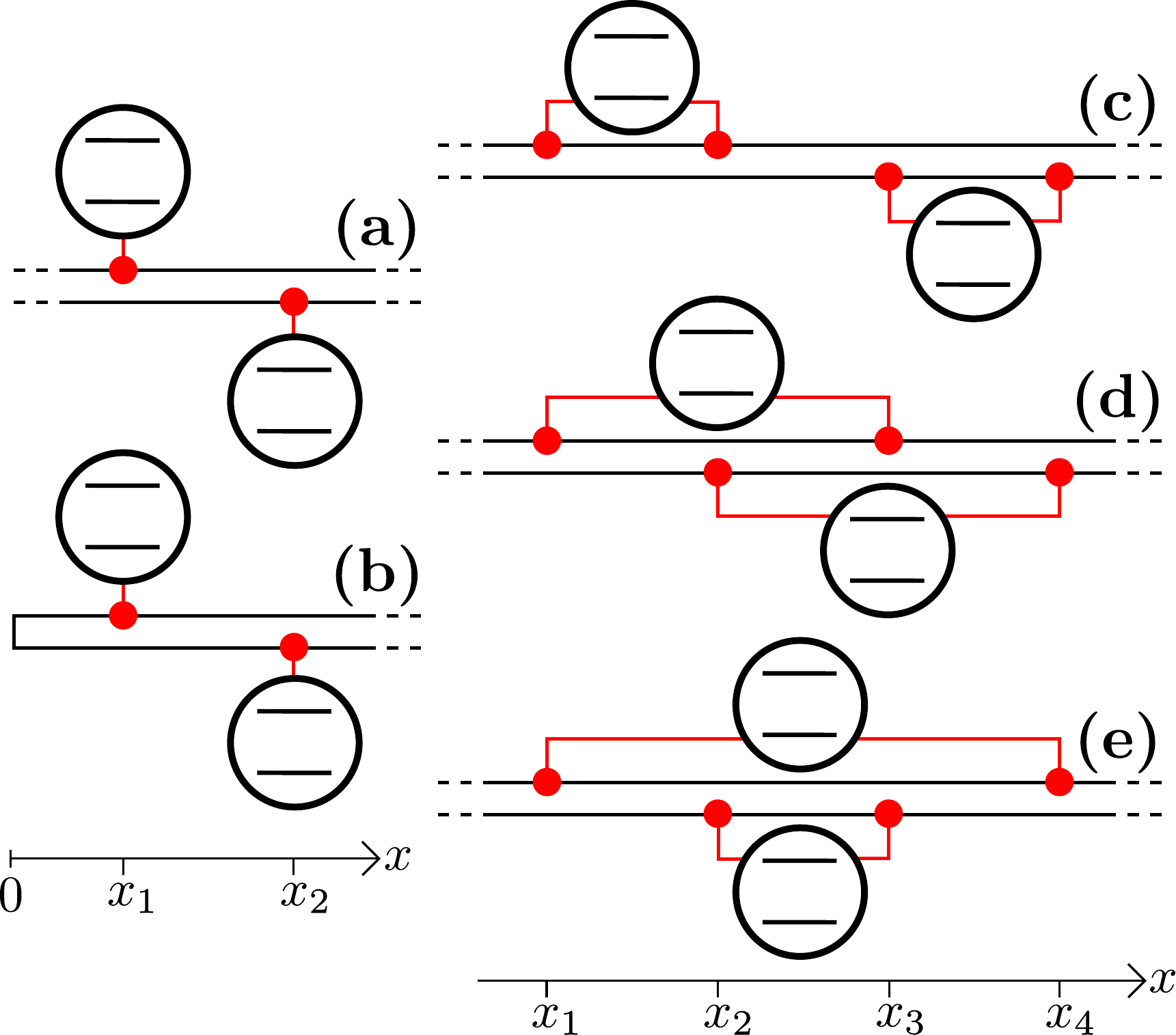}
\caption{Sketches of (a) two small atoms coupled to an open transmission line, (b) two small atoms coupled to a semi-infinite transmission line, (c) two separate giant atoms, (d) two braided giant atoms, and (e) two nested giant atoms. Red circles denote connection points. The atom with the leftmost connection point is denoted $a$ and the other $b$.
\label{fig:2AtomSetups}}
\end{figure}

We begin by comparing setups with two small (i.e., only coupled at a single point) atoms in an open [\figref{fig:2AtomSetups}(a)] or semi-infinite [\figref{fig:2AtomSetups}(b)] waveguide to setups with two giant atoms coupled to an open waveguide at two connection points each. As shown in \figref{fig:2AtomSetups}(c)-(e), there are three distinct topologies for the positions of the connection points in this case. We call the topology in \figref{fig:2AtomSetups}(c) separate giant atoms, the one in \figref{fig:2AtomSetups}(d) braided giant atoms, and the one in \figref{fig:2AtomSetups}(e) nested giant atoms. For simplicity, we limit the discussion in this Letter to atoms with two levels (qubits).

Tracing out the continuum of bosonic modes in the waveguide, a master equation for the density matrix $\rho$ of the atoms can be derived, assuming weak coupling at each connection point and negligible travel time between connection points. We use the SLH formalism~\cite{Gough2009, Gough2009a, Combes2017, Zhang2017} for cascaded quantum systems~\cite{Gardiner1993, Carmichael1993, Gardiner2004} to show~\cite{SupMat} that the master equation for all setups in \figref{fig:2AtomSetups} can be written as
\bea
\dot{\rho} &=& - i \comm{\omega_a' \frac{\sz^a}{2} + \omega_b' \frac{\sz^b}{2} + g \mleft( \sm^a \sp^b + \sp^a \sm^b \mright)}{\rho} \nn\\
&&+ \Gamma_a \lind{\sm^a}\rho +  \Gamma_b \lind{\sm^b}\rho \nn\\
&&+ \Gamma_{\rm coll} \mleft[ \mleft( \sm^a \rho \sp^b - \frac{1}{2}\mleft\{ \sp^a \sm^b , \rho \mright\} \mright) + \text{H.c.} \mright],
\label{eq:ME2Atoms}
\eea
where $\omega_j' = \omega_j + \delta\omega_j$, $\omega_j$ is the transition frequency of atom $j$ only including Lamb shifts from individual connection points, $\delta\omega_j$ is the contribution to the Lamb shift of atom $j$ from interference between connection points, $\lind{A}\rho = A \rho A^\dag - \frac{1}{2}\{A^\dag A, \rho\}$, $g$ is the strength of the exchange interaction between the atoms, $\sp^j$ ($\sm^j$) is the raising (lowering) operator of atom $j$, $\sz^j$ is a Pauli matrix for atom $j$, $\Gamma_j$ is the individual relaxation rate of atom $j$, $\Gamma_{\rm coll}$ is the collective relaxation rate for the atoms, and H.c.~denotes Hermitian conjugate.

\begin{table*}[]
\centering
\renewcommand{\arraystretch}{1.2}
\renewcommand{\tabcolsep}{0.05cm}
\begin{tabular}{ | c | c | c | c | c | c |}
\hline
\textbf{Setup} & \textbf{Frequency shift} $\delta\omega_j$ & \textbf{Exchange interaction} $g$ & \textbf{Individual decay} $\Gamma_j$ & \textbf{Collective decay} $\Gamma_{\rm coll}$ \\
\hline
Small atoms & $0$ & $\frac{\gamma}{2} \sin\varphi$ & $\gamma$ & $\gamma \cos \varphi$ \\ 
\hline
Small atoms + mirror & $\frac{\gamma}{2} \sin \varphi$; $\frac{\gamma}{2} \sin 3\varphi$ & $\frac{\gamma}{2} \mleft( \sin \varphi + \sin 2\varphi \mright)$ & $\gamma\mleft(1 + \cos \varphi \mright)$; $\gamma\mleft(1 + \cos 3\varphi \mright)$ & $\gamma \mleft(\cos \varphi + \cos 2\varphi \mright)$ \\ 
\hline
Separate giant atoms & $\gamma \sin \varphi$ & $\frac{\gamma}{2} \mleft( \sin\varphi + 2\sin 2\varphi + \sin 3\varphi \mright)$ & $2\gamma\mleft(1 + \cos\varphi \mright)$ & $\gamma \mleft(\cos \varphi + 2 \cos 2\varphi + \cos 3\varphi \mright)$ \\
\hline
Braided giant atoms & $\gamma \sin 2\varphi$ & $\frac{\gamma}{2} \mleft( 3 \sin\varphi + \sin 3\varphi \mright)$ & $2\gamma\mleft(1 + \cos 2\varphi \mright)$ & $\gamma \mleft( 3 \cos \varphi + \cos 3\varphi \mright)$ \\ 
\hline
Nested giant atoms & $\gamma \sin 3\varphi$; $\gamma \sin \varphi$ & $\gamma \mleft( \sin\varphi + \sin 2\varphi \mright)$ & $2\gamma\mleft(1 + \cos 3\varphi \mright)$; $2\gamma\mleft(1 + \cos \varphi \mright)$  & $2\gamma \mleft(\cos \varphi + \cos 2\varphi \mright)$ \\
\hline
\end{tabular}
\caption{Frequency shifts, exchange interaction strengths, and decoherence rates for the setups from \figref{fig:2AtomSetups}. In fields with two entries, the first corresponds to atom $a$ and the second to atom $b$.
\label{tab:DecoherenceAndCoupling}}
\end{table*}

For the case of small atoms in an open waveguide, the coefficients in \eqref{eq:ME2Atoms} are already well known~\cite{Lehmberg1970a, Lalumiere2013, Pichler2015a}. In \tabref{tab:DecoherenceAndCoupling}, we compare these coefficients with those that result for the other setups in \figref{fig:2AtomSetups}. For simplicity, we assume here that the distance between subsequent connection points is identical, and that $\omega_a \approx \omega_b$, such that the phase acquired traveling from one connection point to the next is $\varphi = k\abs{x_{j+1}-x_j}$, where the wavenumber $k = \omega_a/ v$, with $v$ the velocity of the modes in the waveguide (for the setup with a mirror, $\varphi = 2 k x_1$). We also assume that the bare relaxation rate at each connection point is $\gamma$. Expressions for arbitrary bare relaxation rates and arbitrary phase shifts between connection points are given in~\cite{SupMat}.

\begin{figure}[]
\centering
\includegraphics[width=\columnwidth]{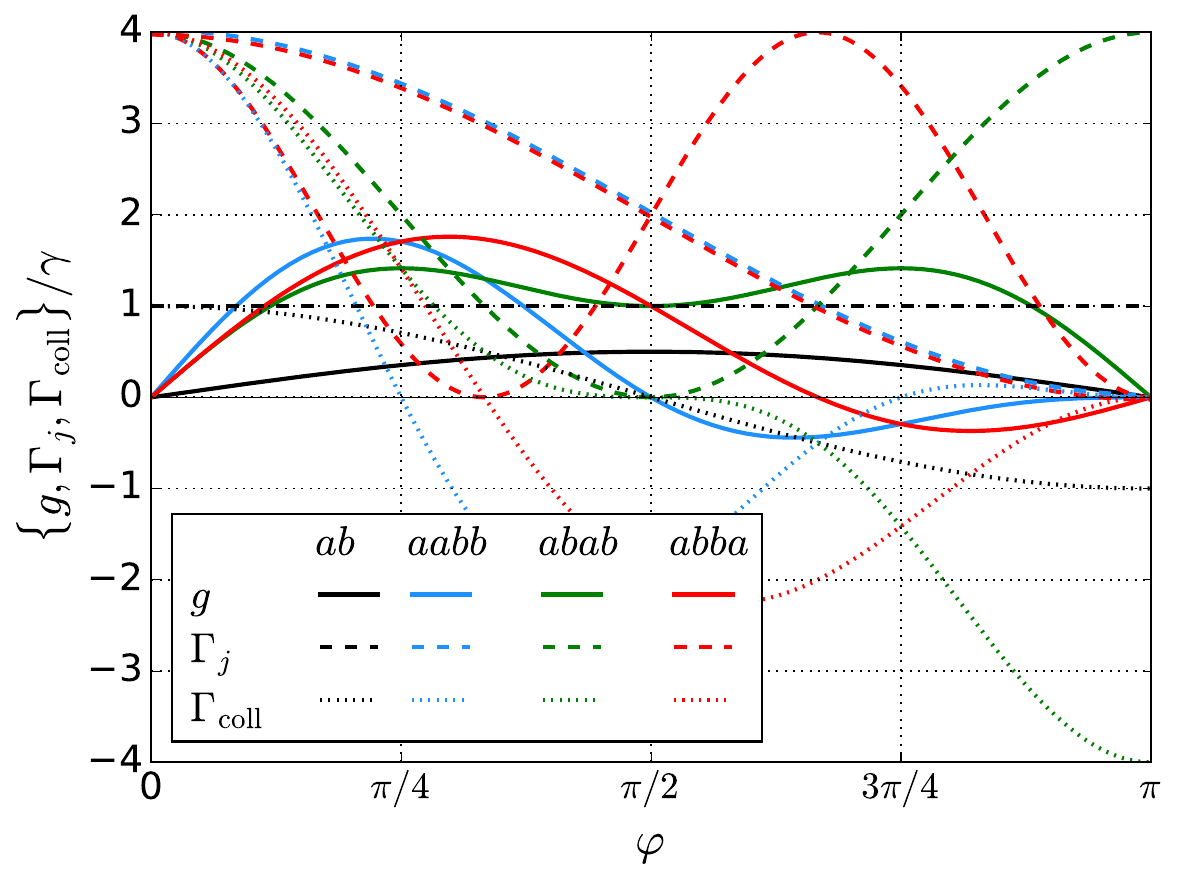}
\caption{Exchange interaction (solid lines) and decoherence rates (individual: dashed lines; collective: dotted lines) as a function of $\varphi$ for the setups in \figref{fig:2AtomSetups}. The corresponding expressions are given in \tabref{tab:DecoherenceAndCoupling}. The labels $ab$ (small atoms, black), $aabb$ (separate giant atoms, blue), $abab$ (braided giant atoms, green), and $abba$ (nested giant atoms, red) correspond to the ordering of connection points for the two atoms. The case of small atoms in a semi-infinite waveguide [\figref{fig:2AtomSetups}(b)] is not plotted separately, since it is qualitatively equivalent to the case of nested giant atoms. Note that there are two red dashed lines, one for $\Gamma_a$ and one for $\Gamma_b$.
\label{fig:DecoherenceAndCouplingComparisons}}
\end{figure}

We plot the relaxation rates and coupling strengths from \tabref{tab:DecoherenceAndCoupling} as functions of $\varphi$ in \figref{fig:DecoherenceAndCouplingComparisons}. For small atoms in an open waveguide, we note that the individual relaxation rates $\Gamma_j \neq 0 \: \forall \varphi$. For this setup, there is only a certain superposition state, the dark state, that is protected from decoherence. For all other setups, there are values of $\varphi$ where $\Gamma_j = 0$. Furthermore, at the points where $\Gamma_j = 0$,  $\Gamma_{\rm coll} = 0$ also holds. Thus, these setups can protect all system states from decoherence. However, the implications of $\Gamma_j = 0$ for $g$ differ for these setups. Only in the case of braided giant atoms is it possible to have $g \neq 0$ when $\Gamma_j = 0 \: \forall j$, i.e., a decoherence-free interaction. This can be understood as follows: $\Gamma_j = 0$ implies that the phase acquired traveling between the connection points of atom $j$ is $(2n + 1) \pi$ for some integer $n$. The collective decay is set by interference between emission from connection points belonging to different atoms, but the sum of these contributions will be zero when the emission from two connection points of one atom interfere destructively. The exchange interaction is set by emission from connection points of one atom being absorbed at connection points of the other atom. For separate and nested giant atoms, the emission from the two connection points belonging to atom $b$ will cancel if $\Gamma_b = 0$, but in the case of \textit{braided} giant atoms, the two inner connection points are placed in-between the two connection points of the other atom, so the contributions from the two connection points of the other atom need \textit{not} interfere destructively. In~\cite{SupMat}, we show that all these conclusions about implications of $\Gamma_j = 0$ for the various setups \textit{remain unchanged} even if we allow for arbitrary bare relaxation rates at each connection point and arbitrary distances (but still negligible travel time) between connection points.


\paragraph*{Generalization to multiple giant atoms with multiple connection points.}

We now consider the most general setup possible, with $N$ atoms such that atom $j$ has $M_j$ connection points and the bare relaxation rate at connection point $j_n$ of atom $j$ is $\gamma_{j_n}$. The phase acquired traveling from connection point $j_n$ of atom $j$ to connection point $k_m$ of atom $k$ is $\varphi_{j_n, k_m}$. With the same assumptions as before, we extend our derivation in the SLH formalism to obtain the master equation~\cite{SupMat}
\bea
&&\dot{\rho} = - i \comm{ \sum_{j = 1}^N \omega_j' \frac{\sz^{(j)}}{2} + \sum_{j = 1}^{N-1} \sum_{k = j + 1}^{N} g_{j,k} \mleft( \sm^{(j)} \sp^{(k)} + \sp^{(j)} \sm^{(k)} \mright)}{\rho} \nn\\
&&+ \sum_{j = 1}^N \Gamma_j \lind{\sm^{(j)}}\rho \nn\\
&&+ \sum_{j = 1}^{N-1} \sum_{k = j + 1}^{N} \Gamma_{{\rm coll}, j, k} \mleft[ \mleft( \sm^{(j)} \rho \sp^{(k)} - \frac{1}{2}\mleft\{ \sp^{(j)} \sm^{(k)}, \rho \mright\} \mright) + \text{H.c.} \mright], \nn\\
\:
\label{eq:MostGeneralME}
\eea
where now $\delta \omega_j = \sum_{n = 1}^{M_j -1} \sum_{m = n +1}^{M_j} \sqrt{\gamma_{j_n} \gamma_{j_m}} \sin \varphi_{j_n, j_m}$, the exchange interaction between atoms $j$ and $k$ is $g_{j,k} =  \sum_{n = 1}^{M_j} \sum_{m = 1}^{M_k} \frac{\sqrt{\gamma_{j_n} \gamma_{k_m}}}{2} \sin \varphi_{j_n, k_m}$, $\Gamma_j = \sum_{n = 1}^{M_j} \sum_{m = 1}^{M_j} \sqrt{\gamma_{j_n} \gamma_{j_m}} \cos \varphi_{j_n, j_m}$, and the collective decay rate for atoms $j$ and $k$ is $ \Gamma_{{\rm coll}, j, k} = \sum_{n = 1}^{M_j} \sum_{m = 1}^{M_k} \sqrt{\gamma_{j_n} \gamma_{k_m}} \cos \varphi_{j_n, k_m}$.

Since all interactions in \eqref{eq:MostGeneralME} are pairwise, the intuition gained from studying the case of two giant atoms with two connection points goes a long way in explaining the properties of these more general setups. If all connection points of atom $j$ are to the left (or right) of all connection points of atom $k$, we call this pair of atoms separate. If all connection points of atom $j$ are situated in-between two subsequent connection points of atom $k$, we call this pair of atoms nested. All other setups are braided. Using the same reasoning as above, we can show that $\Gamma_j = \Gamma_k = 0$ implies both $\Gamma_{{\rm coll}, j, k} = 0$ and $g_{j,k} = 0$ for separate and nested atoms, but $g_{j,k} \neq 0$ is possible if the atoms are braided~\cite{SupMat}.


\paragraph*{1D spin chain with protected, designed nearest-neighbor couplings.}

\begin{figure}[]
\centering
\includegraphics[width=\columnwidth]{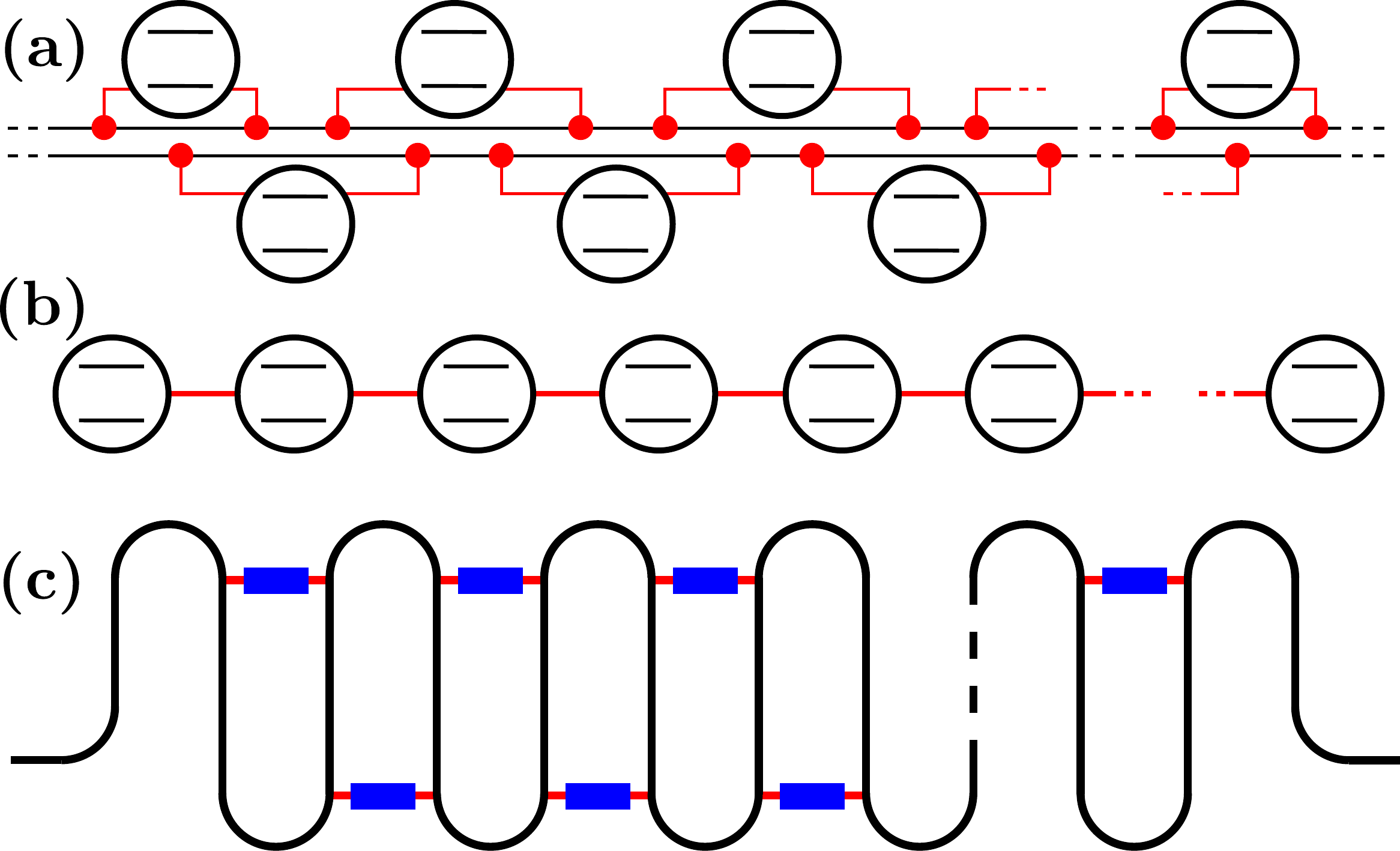}
\caption{Sketch of a setup with giant atoms realizing a 1D chain of qubits with protected nearest-neighbor couplings. (a) The layout of the connection points. (b) The effective system. (c) A possible implementation with superconducting circuits. The black line is a transmission line, the blue blocks are qubits, and the red lines mark where the qubits couple to the transmission line. \label{fig:1DChainSetup}}
\end{figure}

We now discuss two setups with protected pairwise atom-atom interactions that can be realized with multiple giant atoms. The first setup is a 1D chain of atoms with nearest-neighbor couplings, shown in \figref{fig:1DChainSetup}. With the arrangement of connection points given in \figref{fig:1DChainSetup}(a), each pair of neighboring atoms is in a braided configuration, which allows decoherence-free interaction within each such pair, effectively leading to the 1D chain of atoms shown in \figref{fig:1DChainSetup}(b). All other pairs of atoms are \textit{not} braided, and will thus \textit{not} interact when $\Gamma_j = 0 \: \forall j$. In \figref{fig:1DChainSetup}(c), we show how this setup could be implemented with superconducting qubits coupled to a meandering transmission line. Note that there is space for individual readout and control lines to be connected to each qubit in this setup. The decay that such additional channels would introduce can easily be kept small.

If the 1D chain in \figref{fig:1DChainSetup} contains $N$ giant atoms with two connection points each, there will be $2N - 1$ phases between subsequent connection points. Implementing the constraint $\Gamma_j = 0 \: \forall j$ will fix $N$ of these phases. There are then $N-1$ pairwise couplings, set by one phase each: $g_{j, j+1} = \sqrt{\gamma_{(j+1)_1} \gamma_{j_2}} \sin \varphi_{(j+1)_1, j_2}$~\cite{SupMat}. We thus have \textit{maximal freedom} in designing the decoherence-free interactions (both amplitude and sign) in this setup.


\paragraph*{High connectivity for multiple giant atoms.}

\begin{figure}[]
\centering
\includegraphics[width=\columnwidth]{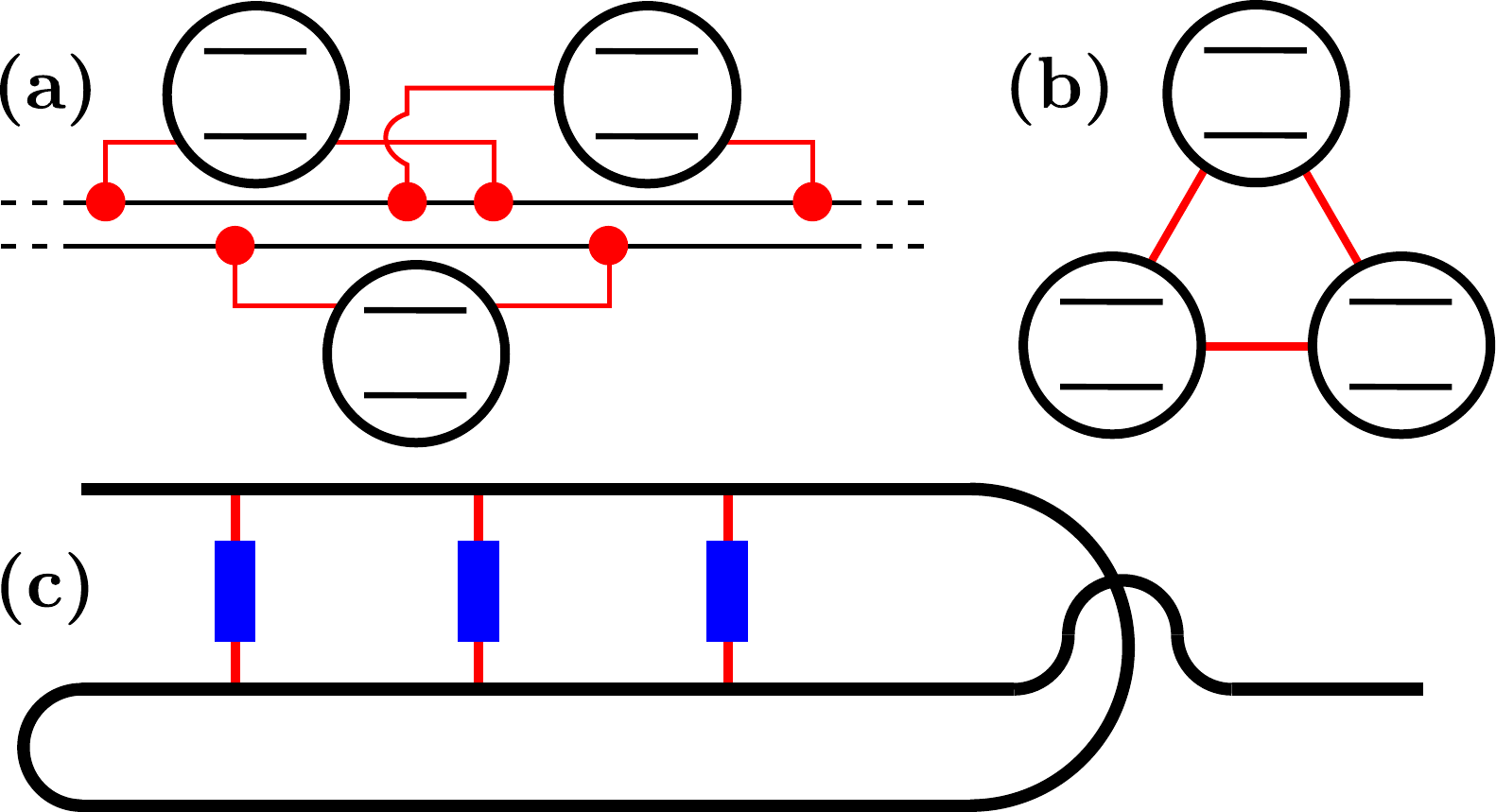}
\caption{Sketch of a setup with three giant atoms realizing protected all-to-all coupling. (a) The layout of the connection points. (b) The effective system. (c) A possible implementation with superconducting circuits. The symbols used are the same as in \figref{fig:1DChainSetup}.
\label{fig:3AtomsSetup}}
\end{figure}

Our second example is a setup with three atoms with a protected all-to-all connectivity, shown in \figref{fig:3AtomsSetup}. With the arrangement of connection points given in \figref{fig:3AtomsSetup}(a), each pair of neighboring atoms is in a braided configuration, which allows decoherence-free interaction within each such pair, effectively leading to the triangular arrangement of coupled atoms shown in \figref{fig:1DChainSetup}(b). In \figref{fig:1DChainSetup}(c), we show how this setup could be implemented with superconducting circuits. Unlike the previous example, this setup requires the transmission line to cross itself at least once, but this can be solved with air bridges~\cite{Riste2015}. Note that it is straightforward to extend this setup to all-to-all connectivity with more atoms by simply adding more superconducting qubits to the row in \figref{fig:1DChainSetup}(c). However, when making $N$ large in the setups in Figs.~\ref{fig:1DChainSetup} and \ref{fig:3AtomsSetup}, care must still be taken that the travel time between connection points remains negligible. This is more important for the setup in \figref{fig:3AtomsSetup} because of the greater connectivity.

For the setup in \figref{fig:3AtomsSetup}, and its generalization to $N$ atoms, the condition $\Gamma_j = 0 \: \forall j$ sets $N$ constraints, which leaves $N-1$ free parameters (phases) to determine the amplitudes of $N(N-1)/2$ pairwise couplings. The individual coupling strengths can thus be chosen quite freely, but not completely freely. In the case $N=3$, we show in~\cite{SupMat} how all couplings can be set to have the same amplitude while their signs can be chosen freely.


\paragraph*{Summary and outlook.}

We have derived a master equation for multiple giant atoms coupled to a 1D waveguide at multiple points, which can be spaced wavelengths apart. We have shown that such giant atoms, with connection points in a braided configuration, can realize a phenomenon that is impossible with small atoms: a \textit{nonzero exchange interaction} mediated by the waveguide between pairs of atoms that are \textit{protected from decoherence} into the waveguide, \textit{regardless} of the atomic state. We have furthermore shown that setups with giant atoms are ready to be implemented in superconducting circuits, and that this could be used for quantum simulation of coupled spins.

This work opens up many interesting directions for future research. Phenomena that have been studied for waveguide QED with small atoms, e.g., sub- and super-radiance (dark and bright states), chiral propagation~\cite{Bliokh2015a, Lodahl2017}, interaction between atoms with more than two levels, and 2D baths~\cite{Gonzalez-Tudela2017a, Gonzalez-Tudela2017b, Perczel2017}, should be revisited to determine whether giant atoms lead to new effects.


\paragraph*{Acknowledgements.}

We acknowledge useful discussions with Lingzhen Guo, Per Delsing, Daniel Campbell, William Oliver, and Simone Gasparinetti. A.F.K.~acknowledges support from a JSPS Postdoctoral Fellowship for Overseas Researchers (P15750). G.J.~acknowledges financial support from the Swedish Research Council and the Knut and Alice Wallenberg Foundation. F.N. acknowledges support from the RIKEN iTHES Project, the MURI Center for Dynamic Magneto-Optics via the AFOSR award number FA9550-14-1-0040, the Japan Society for the Promotion of Science (KAKENHI), the IMPACT program of JST, JSPS-RFBR grant No.~17-52-50023, CREST grant No.~JPMJCR1676, the RIKEN-AIST ``Challenge Research'' program, and the Sir John Templeton Foundation.

\bibliography{GiantAtomRefs}

\end{document}


\title{Supplementary Material for ``Decoherence-free interaction between giant atoms in waveguide QED''}

\author{Anton Frisk Kockum}
\email[e-mail:]{anton.frisk.kockum@gmail.com}
\affiliation{Center for Emergent Matter Science, RIKEN, Saitama 351-0198, Japan}

\author{G\"oran Johansson}
\affiliation{Department of Microtechnology and Nanoscience (MC2), Chalmers University of Technology, SE-412 96 Gothenburg, Sweden}

\author{Franco Nori}
\affiliation{Center for Emergent Matter Science, RIKEN, Saitama 351-0198, Japan}
\affiliation{Physics Department, The University of Michigan, Ann Arbor, Michigan 48109-1040, USA}

\date{\today}

\maketitle

\renewcommand{\thefigure}{S\arabic{figure}}
\renewcommand{\thesection}{S\arabic{section}}
\renewcommand{\theequation}{S\arabic{equation}}
\renewcommand{\bibnumfmt}[1]{[S#1]}
\renewcommand{\citenumfont}[1]{S#1}

\tableofcontents
\vspace{0.3cm}
In this Supplementary Material, we present detailed derivations of all master equations for small and giant atoms used in the main text. We perform these derivations for the most general cases, where both the coupling strengths for different coupling points and the phases acquired between coupling points may differ. We prove the relations between exchange interaction, individual decays, and collective decay in this general setting. We also give input-output relations for multiple giant atoms with multiple connection points.

\section{SLH basics}

We derive all master equations for the multi-atom systems in the SLH formalism \cite{Gough2009, Gough2009a, Combes2017, Zhang2017} for cascaded quantum systems. To make the treatment here self-contained, we first state the basic properties and rules of this formalism. For more details, see the review in Ref.~\cite{Combes2017}.

An open quantum system with $n$ input-output ports can be described by an SLH triplet $G = \mleft(\mathbf{S}, \mathbf{L}, H \mright)$, where $\mathbf{S}$ is an $n \times n$ scattering matrix, $\mathbf{L}$ is an $n \times 1$ vector describing the coupling of the system to the environment at the input-output ports (e.g., if a cavity with annihilation operator $a$ is leaking photons at a rate $\kappa$ through one of its mirrors, and this mirror constitutes the $j$th input-output port of the system, that will give an entry $L_j = \sqrt{\kappa} a$ in $\mathbf{L}$), and $H$ is the Hamiltonian of the system.

\begin{figure}
\centering
\includegraphics[width=0.8\linewidth]{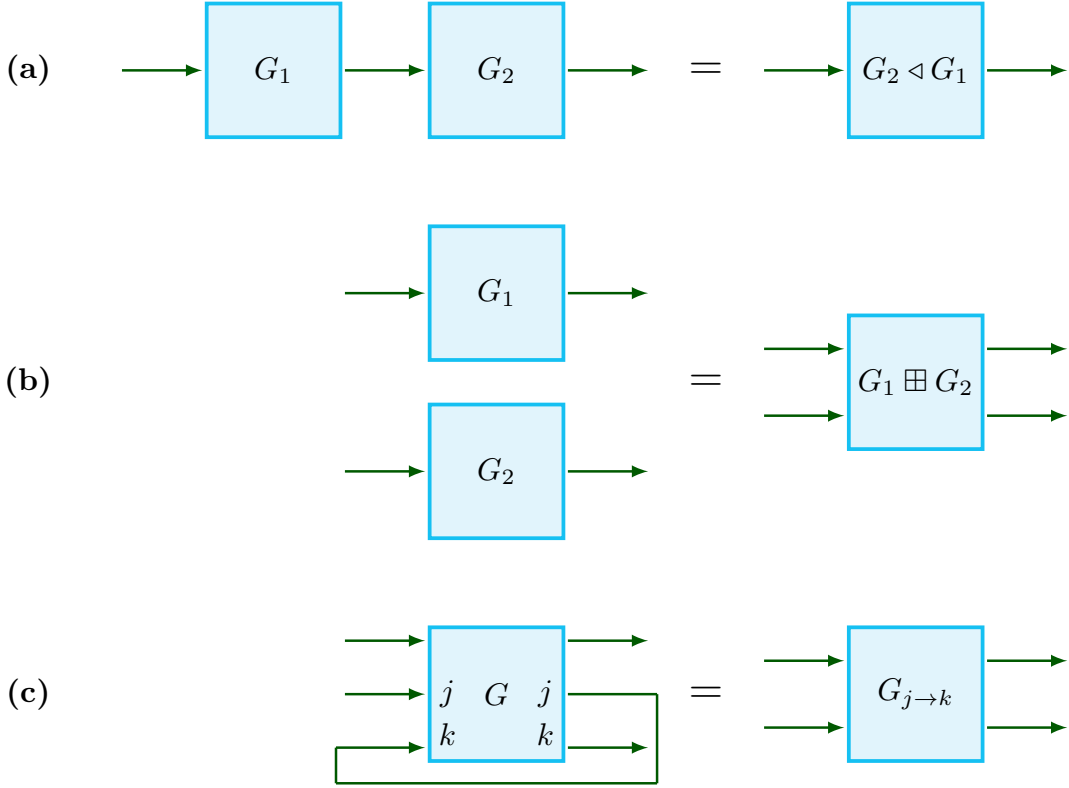}
\caption{Illustrations of the operations covered by the three basic composition rules in the SLH formalism. (a) The series product $G_2 \triangleleft G_1$. (b) The concatenation product $G_1 \boxplus G_2$. (c) The feedback operation $G_{j \rightarrow k}$. \label{fig:SLHRules}}
\end{figure}

To combine SLH triplets of cascaded quantum systems into a single triplet describing the entire setup, three composition rules, illustrated in \figref{fig:SLHRules}, are used: the series product $\triangleleft$, the concatenation product $\boxplus$, and a feedback operation. The series product, shown in \figref{fig:SLHRules}(a), is used when the outputs from a system described by an SLH triplet $G_1$ are used as inputs to another system, described by an SLH triplet $G_2$ with the same number of input-output ports, such that the $j$th output from the first system becomes the $j$th input to the second system. The resulting total SLH triplet is then
\be
G_2 \triangleleft G_1 = \mleft( \mathbf{S}_2 \mathbf{S}_1, \mathbf{S}_2 \mathbf{L}_1 + \mathbf{L}_2, H_1 + H_2 + \frac{1}{2i} \mleft[ \mathbf{L}_2^\dag \mathbf{S}_2 \mathbf{L}_1 - \mathbf{L}_1^\dag \mathbf{S}_2^\dag \mathbf{L}_2 \mright] \mright).
\label{eq:SeriesProduct}
\ee
If two systems, described by $G_1$ and $G_2$, are combined in parallel, the resulting total SLH triplet is given by the concatenation product
\be
G_1 \boxplus G_2 = \mleft( \begin{bmatrix} \mathbf{S}_1 & 0 \\ 0 & \mathbf{S}_2 \end{bmatrix}, \begin{bmatrix} \mathbf{L}_1 \\  \mathbf{L}_2 \end{bmatrix}, H_1 + H_2 \mright), 
\label{EqConcatenationProduct}
\ee
shown in \figref{fig:SLHRules}(b). Finally, if the $j$th output of a system described by the triplet $G = (\mathbf{S}, \mathbf{L}, H)$ is fed back as the $k$th input of the same system, as illustrated in \figref{fig:SLHRules}(c), the rule for feedback reduction states that the resulting system is described by the triplet $G_{j \rightarrow k} = (\tilde{\mathbf{S}}, \tilde{\mathbf{L}}, \tilde{H})$, where
\bea
\tilde{\mathbf{S}} &=& \mathbf{S}_{\bar{j} \bar{k}} + \mathbf{S}_{\bar{j} k} \mleft( 1 - S_{jk} \mright)^{-1} \mathbf{S}_{j \bar{k}}, \\
\tilde{\mathbf{L}} &=& \mathbf{L}_{\bar{j}} + \mathbf{S}_{\bar{j} k} \mleft( 1 - S_{jk} \mright)^{-1} L_j, \\
\tilde{H} &=& H + \frac{1}{2i} \mleft[ \mathbf{L}^\dag \mathbf{S}_{: k} \mleft( 1 - S_{jk} \mright)^{-1} L_j - \text{H.c.} \mright].
\eea
Here, $\mathbf{S}_{\bar{j} \bar{k}}$ denotes $\mathbf{S}$ with the $j$th row and $k$th column removed, $\mathbf{S}_{\bar{j} k}$
denotes the $k$th column of $\mathbf{S}$ with the $j$th row removed, $\mathbf{S}_{j \bar{k}}$ denotes the $j$th row of $\mathbf{S}$ with the $k$th column removed, $S_{jk}$ is the element in the $j$th row and $k$th column of $\mathbf{S}$, $\mathbf{S}_{: k}$ is the $k$th column of $\mathbf{S}$, and H.c.~denotes Hermitian conjugate.

When the SLH triplet $G = (\mathbf{S}, \mathbf{L}, H)$ for a system has been found, the master equation for the system is given by
\be
\dot{\rho} = -i\comm{H}{\rho} + \sum_{j=1}^n \lind{L_j}\rho,
\label{eq:SLHGeneralME}
\ee
where $\lind{X}\rho = X\rho X^\dag - \frac{1}{2} X^\dag X\rho - \frac{1}{2} \rho X^\dag X$ are Lindblad operators. The output from port $j$ of the system is simply given by $L_j$. 

Note that the SLH formalism relies on the same physical assumptions as the standard Lindblad master equation, i.e., weak coupling and the Markov approximation. In addition, the SLH formalism as presented so far also requires the fields connecting various systems to propagate in linear, dispersion-less media, and that the propagation time is negligible. Furthermore, it is also assumed that all input fields are in the vacuum state, but non-vacuum inputs can be incorporated by introducing triplets for various sources. For example, if input port $j$ of a system with $N$ input-output ports and triplet $G = (\mathbf{S}, \mathbf{L}, H)$ is driven by a coherent drive supplying $\abssq{\alpha}$ photons per second, this can be modeled by $ G_1 \triangleleft \mleft( \mathbbm{I}_{j-1} \boxplus G_{\alpha} \boxplus \mathbbm{I}_{N-j} \mright)$, where $\mathbbm{I}_k = \mleft(\mathbf{1}_k, \mathbf{0}, 0 \mright)$, $\mathbf{1}_k$ is the $k \times k$ identity matrix, and $G_{\alpha} = \mleft( 1, \alpha, 0 \mright)$, in the rotating frame of the drive.


\section{Master equations for two small atoms}

As a prelude to the calculations for giant atoms, we first re-derive the known master equation for two small atoms in an open waveguide using the SLH formalism. We then also study the case of two small atoms in a semi-infinite waveguide, which has similarities with giant-atom setups.


\subsection{Open waveguide}
\label{sec:SupMatME2SmallAtoms}

\begin{figure}
\centering
\includegraphics[width=0.9\linewidth]{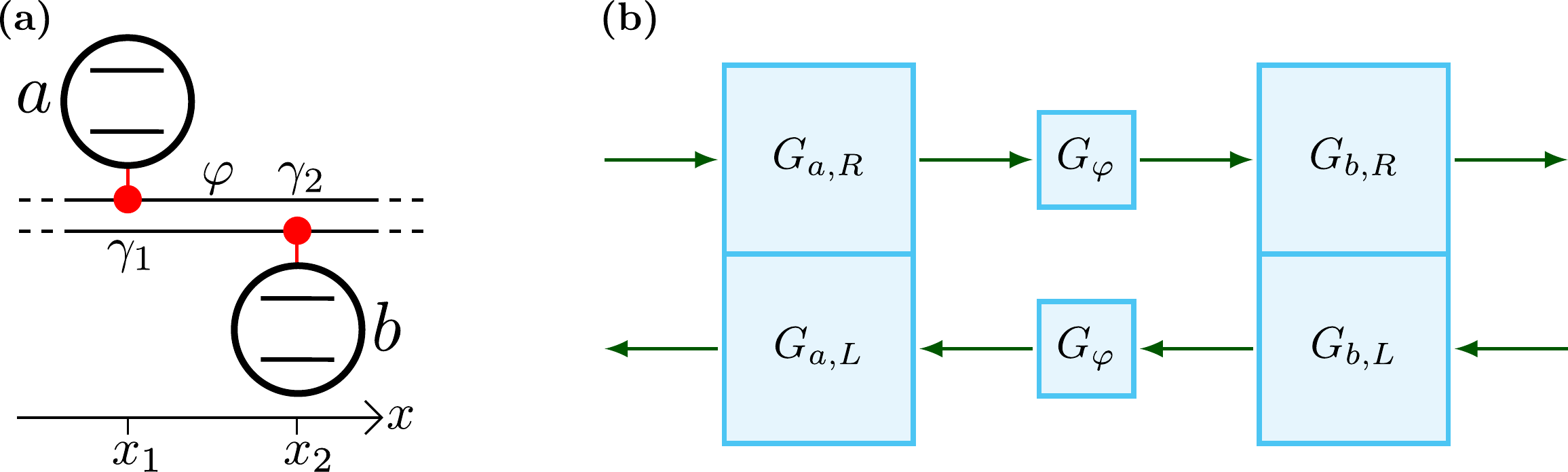}
\caption{Two small atoms in an open waveguide. (a) A sketch showing the relevant parameters. (b) The input-output flows in the corresponding SLH calculation. \label{fig:2SmallAtomsSketchAndSLH}}
\end{figure}

We first consider the setup shown in \figref{fig:2SmallAtomsSketchAndSLH}(a), where the two atoms, with resonance frequencies $\omega_a$ and $\omega_b$, respectively, have relaxation rates $\gamma_1$ and $\gamma_2$, respectively, due to their coupling to the waveguide. The distance between the atoms is such that a signal propagating in the waveguide acquires a phase shift $\varphi$ when traveling between them. This phase shift is calculated as $\varphi = k \abs{x_2 - x_1}$, where $k = \omega_k / v$ is the wavenumber given by an angular frequency $\omega_k$ and the propagation velocity in the waveguide, $v$. We assume $\omega_k = \omega_a \approx \omega_b$. This is consistent with the assumptions behind the SLH formalism. As long as the relaxation rates $\gamma_j$ are small compared to the transition frequencies $\omega_j$, and the distance $\abs{x_2 - x_1}$ is not much more than a wavelength, only frequencies in a bandwidth of about $\gamma_j$ will contribute to the dynamics. This bandwidth is small enough that variations in the phase shift, due to different frequencies within this bandwidth, are negligible. If $\omega_a$ and $\omega_b$ differ by more than $\gamma_j$, the exchange interaction between atoms $a$ and $b$ will be negligible in the rotating-wave approximation.

There are both right- and left-propagating modes in the waveguide. To handle this in the SLH formalism, the easiest approach is to calculate the cascading operations separately for these modes and then concatenate the results. Explicitly, we divide the system into SLH triplets as shown in \figref{fig:2SmallAtomsSketchAndSLH}(b), separating the triplets for the atoms into two parts with the Hamiltonians residing in the part coupling to the right-propagating modes (we use $\hbar = 1$ throughout this article):
\bea
G_a &=& \mleft( \begin{bmatrix} 1 & 0 \\ 0 & 1 \end{bmatrix}, \begin{bmatrix} \sqrt{\frac{\gamma_1}{2}}\sm^a \\ \sqrt{\frac{\gamma_1}{2}}\sm^a \end{bmatrix} , \omega_a \frac{\sz^a}{2} \mright) = G_{a,\rm R} \boxplus G_{a,\rm L} = \mleft( 1, \sqrt{\frac{\gamma_1}{2}}\sm^a, \omega_a \frac{\sz^a}{2} \mright) \boxplus \mleft( 1, \sqrt{\frac{\gamma_1}{2}}\sm^a, 0 \mright),
\label{eq:Ga2SmallAtoms} \\
G_b &=& \mleft( \begin{bmatrix} 1 & 0 \\ 0 & 1 \end{bmatrix}, \begin{bmatrix} \sqrt{\frac{\gamma_2}{2}}\sm^b \\ \sqrt{\frac{\gamma_2}{2}}\sm^b \end{bmatrix} , \omega_b \frac{\sz^b}{2} \mright) = G_{b,\rm R} \boxplus G_{b,\rm L} = \mleft( 1, \sqrt{\frac{\gamma_2}{2}}\sm^b, \omega_b \frac{\sz^b}{2} \mright) \boxplus \mleft( 1, \sqrt{\frac{\gamma_2}{2}}\sm^b, 0 \mright).
\label{eq:Gb2SmallAtoms}
\eea
The phase shift is included through the triplet $G_\varphi = \mleft( e^{i\varphi}, 0, 0 \mright)$. From the rule for the series product given in \eqref{eq:SeriesProduct} we obtain the triplet for the right-propagating modes
\bea
G_{\rm R} &=& G_{b, \rm R} \triangleleft G_\varphi \triangleleft G_{a, \rm R} = \mleft( e^{i\varphi}, \sqrt{\frac{\gamma_2}{2}}\sm^b, \omega_b \frac{\sz^b}{2} \mright) \triangleleft G_{a, \rm R} \nn\\
&=& \mleft( e^{i\varphi}, e^{i\varphi} \sqrt{\frac{\gamma_1}{2}}\sm^a + \sqrt{\frac{\gamma_2}{2}}\sm^b, \omega_a \frac{\sz^a}{2} + \omega_b \frac{\sz^b}{2} + \frac{\sqrt{\gamma_1 \gamma_2}}{4i} \mleft[ e^{i\varphi} \sm^a \sp^b - e^{-i\varphi} \sp^a \sm^b \mright] \mright),
\label{eq:GR2SmallAtoms}
\eea
and, in the same way, the triplet for the left-propagating modes
\bea
G_{\rm L} &=& G_{a, \rm L} \triangleleft G_\varphi \triangleleft G_{b, \rm L} = \mleft( e^{i\varphi}, \sqrt{\frac{\gamma_1}{2}}\sm^a + e^{i\varphi} \sqrt{\frac{\gamma_2}{2}}\sm^b, \frac{\sqrt{\gamma_1 \gamma_2}}{4i} \mleft[ e^{i\varphi} \sp^a \sm^b - e^{-i\varphi} \sm^b \sp^a \mright] \mright).
\label{eq:GL2SmallAtoms}
\eea
Concatenating these triplets, the final result is
\bea
G_{\rm tot} = G_{\rm R} \boxplus G_{\rm L} = \mleft( \begin{bmatrix} e^{i\varphi} & 0 \\ 0 & e^{i\varphi} \end{bmatrix}, \begin{bmatrix} e^{i\varphi} \sqrt{\frac{\gamma_1}{2}}\sm^a + \sqrt{\frac{\gamma_2}{2}}\sm^b \\ \sqrt{\frac{\gamma_1}{2}}\sm^a + e^{i\varphi} \sqrt{\frac{\gamma_2}{2}}\sm^b \end{bmatrix}, \omega_a \frac{\sz^a}{2} + \omega_b \frac{\sz^b}{2} + \frac{\sqrt{\gamma_1 \gamma_2}}{2} \sin \varphi \mleft[ \sm^a \sp^b + \sp^a \sm^b \mright] \mright). \qquad
\label{eq:Gtot2AtomsOpenWaveguide}
\eea

From the SLH triplet $G_{\rm tot}$ in \eqref{eq:Gtot2AtomsOpenWaveguide}, we can extract the master equation for the system using \eqref{eq:SLHGeneralME}. In this calculation, we use the following property of the Lindblad operators:
\be
\lind{a + b}\rho = \lind{a}\rho + \lind{b}\rho + a\rho b^\dag + b\rho a^\dag - \frac{1}{2}\mleft[ \mleft(a^\dag b + b^\dag a \mright) \rho + \rho \mleft( a^\dag b + b^\dag a \mright) \mright].
\label{eq:Dab}\\
\ee
The resulting master equation is
\bea
\dot{\rho} &=& - i \comm{\omega_a \frac{\sz^a}{2} + \omega_b \frac{\sz^b}{2} + \frac{\sqrt{\gamma_1 \gamma_2}}{2} \sin \varphi \mleft( \sm^a \sp^b + \sp^a \sm^b \mright)}{\rho} \nn\\
&&+ \lind{e^{i\varphi} \sqrt{\frac{\gamma_1}{2}}\sm^a + \sqrt{\frac{\gamma_2}{2}}\sm^b}\rho + \lind{\sqrt{\frac{\gamma_1}{2}}\sm^a + e^{i\varphi} \sqrt{\frac{\gamma_2}{2}}\sm^b}\rho \nn\\
&=& - i \comm{\omega_a \frac{\sz^a}{2} + \omega_b \frac{\sz^b}{2} + \frac{\sqrt{\gamma_1 \gamma_2}}{2} \sin \varphi \mleft( \sm^a \sp^b + \sp^a \sm^b \mright)}{\rho} \nn\\
&&+ \gamma_1 \lind{\sm^a}\rho + \gamma_2 \lind{\sm^b}\rho + \sqrt{\gamma_1 \gamma_2} \cos \varphi \mleft\{ \sm^a \rho \sp^b + \sm^b \rho \sp^a - \frac{1}{2}\mleft[ \mleft(\sp^a \sm^b + \sp^b \sm^a \mright) \rho + \rho \mleft( \sp^a \sm^b + \sp^b \sm^a \mright) \mright] \mright\}. \qquad
\label{eq:ME2AtomsOpenWaveguide}
\eea
Assuming equal relaxation rates for the two atoms, i.e., setting $\gamma_1 = \gamma_2 \equiv \gamma$, we see that the master equation in \eqref{eq:ME2AtomsOpenWaveguide} reduces to Eq.~(1) of the main text with the coefficients given in the second row of Table I.


\subsection{Semi-infinite waveguide}

\begin{figure}
\centering
\includegraphics[width=0.9\linewidth]{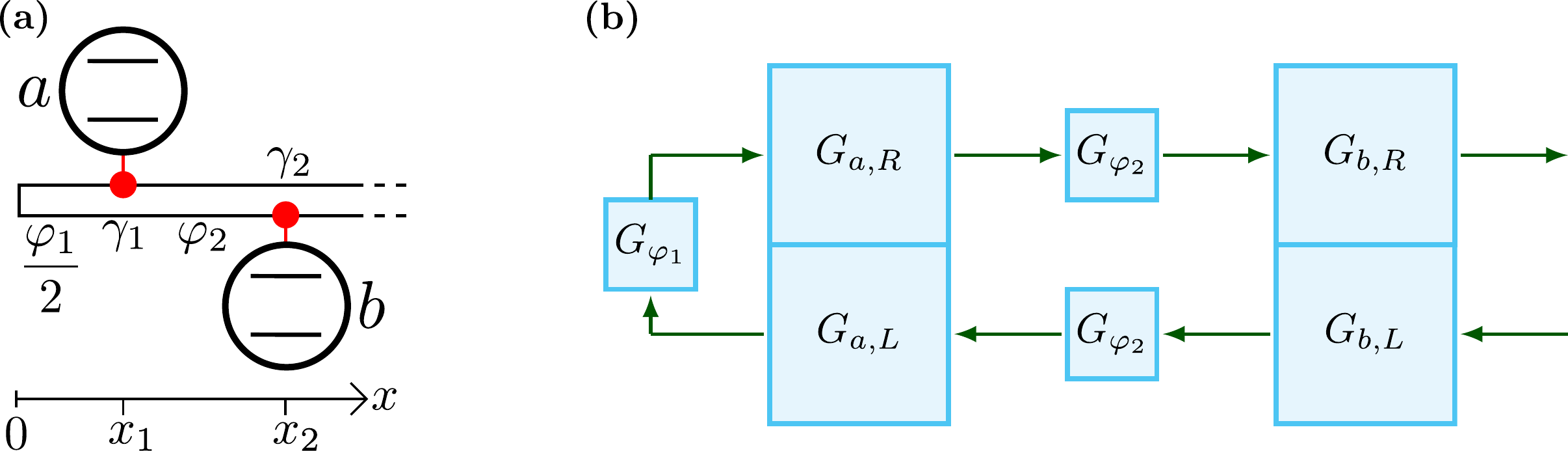}
\caption{Two small atoms in a semi-infinite waveguide. (a) A sketch showing the relevant parameters. (b) The input-output flows in the corresponding SLH calculation. \label{fig:2SmallAtomsMirrorSketchAndSLH}}
\end{figure}

Adding a mirror to make the open waveguide, to which the two small atoms couple, semi-infinite instead, we have the setup shown in \figref{fig:2SmallAtomsMirrorSketchAndSLH}(a). The phase acquired when traveling from atom $a$ to the mirror is $\varphi_1/2$, so the phase acquired during a roundtrip from atom $a$ to the mirror and back is $\varphi_1$.

To calculate the SLH triplet for this setup, we again start by separating the triplets for the atoms into parts interacting with the left- and right-propagating modes in the transmission line, exactly like in Eqs.~(\ref{eq:Ga2SmallAtoms})-(\ref{eq:Gb2SmallAtoms}). All parts needed for the SLH calculation are shown in \figref{fig:2SmallAtomsMirrorSketchAndSLH}(b). Unlike in the open-waveguide case, here we do not use the concatenation product to add up the left- and right-moving parts; instead, we combine through the series product due to the presence of the mirror. The triplet for the whole system is
\be
G_{\rm tot} = G_{b, \rm R} \triangleleft G_{\varphi_2} \triangleleft G_{a, \rm R} \triangleleft G_{\varphi_1} \triangleleft G_{a, \rm L} \triangleleft G_{\varphi_2} \triangleleft G_{b, \rm L}
\ee
Using the results in Eqs.~(\ref{eq:GR2SmallAtoms})-(\ref{eq:GL2SmallAtoms}) leads to
\bea
G_{\rm tot} &=& \mleft( e^{i\varphi_2}, e^{i\varphi_2} \sqrt{\frac{\gamma_1}{2}}\sm^a + \sqrt{\frac{\gamma_2}{2}}\sm^b, \omega_a \frac{\sz^a}{2} + \omega_b \frac{\sz^b}{2} + \frac{\sqrt{\gamma_1 \gamma_2}}{4i} \mleft[ e^{i\varphi_2} \sm^a \sp^b - e^{-i\varphi_2} \sp^a \sm^b \mright] \mright) \triangleleft G_{\varphi_1} \nn\\
&& \triangleleft \mleft( e^{i\varphi_2}, \sqrt{\frac{\gamma_1}{2}}\sm^a + e^{i\varphi_2} \sqrt{\frac{\gamma_1}{2}}\sm^b, \frac{\sqrt{\gamma_1 \gamma_2}}{4i} \mleft[ e^{i\varphi_2} \sp^a \sm^b - e^{-i\varphi_2} \sm^b \sp^a \mright] \mright)
\eea
By repeated application of the series-product rule, we arrive at the three components of $G_{\rm tot}$:
\bea
\mathbf{S}_{\rm tot} &=& e^{i \mleft( \varphi_1 + 2 \varphi_2 \mright)}, \\
\mathbf{L}_{\rm tot} &=& e^{i \varphi_2} \mleft( 1 + e^{i \varphi_1} \mright) \sqrt{\frac{\gamma_1}{2}} \sm^a + \mleft( 1 + e^{i \mleft( \varphi_1 + 2 \varphi_2 \mright)} \mright) \sqrt{\frac{\gamma_2}{2}} \sm^b, \\
H_{\rm tot} &=& \mleft(\omega_a + \frac{\gamma_1}{2} \sin \varphi_1 \mright) \frac{\sz^a}{2} + \mleft[ \omega_b + \frac{\gamma_2}{2} \sin \mleft( \varphi_1 + 2 \varphi_2 \mright) \mright] \frac{\sz^b}{2} + \frac{\sqrt{\gamma_1 \gamma_2}}{2} \mleft[ \sin \varphi_2 + \sin \mleft( \varphi_1 + \varphi_2 \mright) \mright] \mleft( \sm^a \sp^b + \sp^a \sm^b \mright).
\eea
To obtain the final form of $H_{\rm tot}$, we used the identity $\sp \sm = \mleft( 1 + \sz \mright) / 2$ and the fact that constant terms can be excluded from the Hamiltonian since they do not contribute to the dynamics.

With the total triplet in hand, we extract the master equation in the same way as for the open-waveguide case:
\bea
\dot{\rho} &=& - i \comm{H_{\rm tot}}{\rho} + \lind{e^{i \varphi_2} \mleft( 1 + e^{i \varphi_1} \mright) \sqrt{\frac{\gamma_1}{2}} \sm^a + \mleft( 1 + e^{i \mleft( \varphi_1 + 2 \varphi_2 \mright)} \mright) \sqrt{\frac{\gamma_2}{2}} \sm^b} \rho \nn\\
&=& - i \comm{H_{\rm tot}}{\rho} + \gamma_1 \mleft( 1 + \cos \varphi_1 \mright) \lind{\sm^a} \rho + \gamma_2 \mleft[ 1 + \cos \mleft( \varphi_1 + 2 \varphi_2 \mright) \mright] \lind{\sm^a} \rho \nn\\
&&+ \sqrt{\gamma_1 \gamma_2} \mleft[ \cos \varphi_2 + \cos \mleft( \varphi_1 + \varphi_2 \mright) \mright] \mleft\{ \sm^a \rho \sp^b + \sm^b \rho \sp^a - \frac{1}{2}\mleft[ \mleft(\sp^a \sm^b + \sp^b \sm^a \mright) \rho + \rho \mleft( \sp^a \sm^b + \sp^b \sm^a \mright) \mright] \mright\}.
\label{eq:ME2AtomsMirror}
\eea
If we assume equal relaxation rates for the two atoms ($\gamma_1 = \gamma_2 \equiv \gamma$) and equal phases ($\varphi_1 = \varphi_2 \equiv \varphi$), \eqref{eq:ME2AtomsMirror} reduces to Eq.~(1) of the main text with the coefficients given in the third row of Table I.


\section{Master equations for two giant atoms with two connection points}

We now use the SLH formalism to derive the master equations for all geometries with giant atoms coupled to an open waveguide at two connection points. The remarks about approximations made for the phase shift for two small atoms in the preceding section are valid here as well, and also for more than two giant atoms with more than two connection points.


\subsection{Separate giant atoms}

\begin{figure}
\centering
\includegraphics[width=\linewidth]{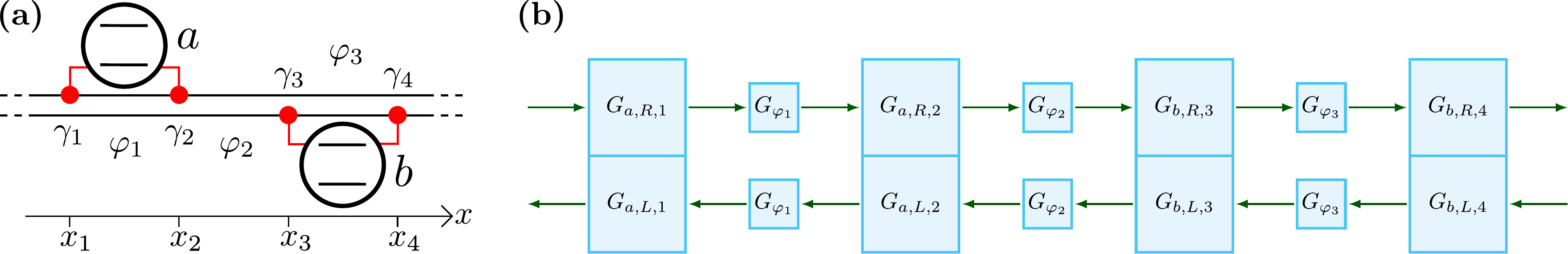}
\caption{Two separate giant atoms in an open waveguide. (a) A sketch showing the relevant parameters. (b) The input-output flows in the corresponding SLH calculation. \label{fig:2SeparateGiantAtomsSketchAndSLH}}
\end{figure}

We first consider two separate giant atoms. The setup, with definitions of all coupling strengths and phase shifts, is shown in \figref{fig:2SeparateGiantAtomsSketchAndSLH}(a). To calculate the SLH triplet for this setup, we follow the scheme sketched in \figref{fig:2SeparateGiantAtomsSketchAndSLH}(b). The triplets for the atoms are first decomposed into the following parts:
\bea
G_{a, \rm R, 1} &=& \mleft(1, \sqrt{\frac{\gamma_1}{2}}\sm^a, \omega_a \frac{\sz^a}{2} \mright), \\
G_{a, \rm R, 2} &=& \mleft(1, \sqrt{\frac{\gamma_2}{2}}\sm^a, 0 \mright), \\
G_{a, \rm L, 1} &=& \mleft(1, \sqrt{\frac{\gamma_1}{2}}\sm^a, 0 \mright), \\
G_{a, \rm L, 2} &=& \mleft(1, \sqrt{\frac{\gamma_2}{2}}\sm^a, 0 \mright), \\
G_{b, \rm R, 3} &=& \mleft(1, \sqrt{\frac{\gamma_3}{2}}\sm^b, \omega_b \frac{\sz^b}{2} \mright), \\
G_{b, \rm R, 4} &=& \mleft(1, \sqrt{\frac{\gamma_4}{2}}\sm^b, 0 \mright), \\
G_{b, \rm L, 3} &=& \mleft(1, \sqrt{\frac{\gamma_3}{2}}\sm^b, 0 \mright), \\
G_{b, \rm L, 4} &=& \mleft(1, \sqrt{\frac{\gamma_4}{2}}\sm^b, 0 \mright).
\eea
The triplet for the right-moving part is then given by
\be
G_{\rm R} = G_{b, \rm R, 4} \triangleleft G_{\varphi_3} \triangleleft G_{b, \rm R, 3} \triangleleft G_{\varphi_2} \triangleleft G_{a, \rm R, 2} \triangleleft G_{\varphi_1} \triangleleft G_{a, \rm R, 1}.
\ee
Using the series product rule, the first part of this expression becomes
\be
G_{b, \rm R, 4} \triangleleft G_{\varphi_3} \triangleleft G_{b, \rm R, 3} = \mleft( e^{i \varphi_3}, \mleft[ e^{i \varphi_3} \sqrt{\frac{\gamma_3}{2}} + \sqrt{\frac{\gamma_4}{2}} \mright] \sm^b, \mleft( \omega_b + \frac{\sqrt{\gamma_3 \gamma_4}}{2} \sin \varphi_3 \mright) \frac{\sz^b}{2} \mright),
\ee
where we again removed a constant term in the Hamiltonian. From symmetry, we then immediately obtain
\be
G_{a, \rm R, 1} \triangleleft G_{\varphi_1} \triangleleft G_{a, \rm R, 2} = \mleft( e^{i \varphi_1}, \mleft[ e^{i \varphi_1} \sqrt{\frac{\gamma_1}{2}} + \sqrt{\frac{\gamma_2}{2}} \mright] \sm^a, \mleft( \omega_a + \frac{\sqrt{\gamma_1 \gamma_2}}{2} \sin \varphi_1 \mright) \frac{\sz^a}{2} \mright).
\ee
These results, together with repeated application of the series product rule, lead to $G_{\rm R} = \mleft(\mathbf{S}_{\rm R}, \mathbf{L}_{\rm R}, H_{\rm R} \mright)$ with
\bea
\mathbf{S}_{\rm R} &=& e^{i \mleft( \varphi_1 + \varphi_2 + \varphi_3 \mright)}, \\
\mathbf{L}_{\rm R} &=& \mleft( e^{i \mleft( \varphi_1 + \varphi_2 + \varphi_3 \mright)} \sqrt{\frac{\gamma_1}{2}} + e^{i \mleft( \varphi_2 + \varphi_3 \mright)} \sqrt{\frac{\gamma_2}{2}} \mright) \sm^a + \mleft( e^{i \varphi_3} \sqrt{\frac{\gamma_3}{2}} + \sqrt{\frac{\gamma_4}{2}} \mright) \sm^b, \\
H_{\rm R} &=& \mleft( \omega_a + \frac{\sqrt{\gamma_1 \gamma_2}}{2} \sin \varphi_1 \mright) \frac{\sz^a}{2} + \mleft( \omega_b + \frac{\sqrt{\gamma_3 \gamma_4}}{2} \sin \varphi_3 \mright) \frac{\sz^b}{2} \nn\\
&&+ \frac{1}{4i} \mleft[ \mleft( e^{i \mleft( \varphi_1 + \varphi_2 + \varphi_3 \mright)} \sqrt{\gamma_1 \gamma_4} + e^{i \mleft( \varphi_2 + \varphi_3 \mright)} \sqrt{\gamma_2 \gamma_4} + e^{i \mleft( \varphi_1 + \varphi_2 \mright)} \sqrt{\gamma_1 \gamma_3} + e^{i \varphi_2} \sqrt{\gamma_2 \gamma_3} \mright) \sm^a \sp^b - \text{H.c.} \mright].
\eea
From the symmetry that is apparent in \figref{fig:2SeparateGiantAtomsSketchAndSLH}(b), we can immediately deduce that the triplet for the left-moving part, $G_{\rm L} = \mleft(\mathbf{S}_{\rm L}, \mathbf{L}_{\rm L}, H_{\rm L} \mright)$, is given by removing the $\omega_j$ parts in the Hamiltonian, making the substitution $\varphi_1 \leftrightarrow \varphi_3$, and changing the non-$\varphi$ indices according to $a \leftrightarrow b$, $1 \leftrightarrow 4$ and $2 \leftrightarrow 3$ in the equations for $G_{\rm R}$, i.e.,
\bea
\mathbf{S}_{\rm L} &=& e^{i \mleft( \varphi_1 + \varphi_2 + \varphi_3 \mright)}, \\
\mathbf{L}_{\rm L} &=& \mleft( e^{i \mleft( \varphi_1 + \varphi_2 + \varphi_3 \mright)} \sqrt{\frac{\gamma_4}{2}} + e^{i \mleft( \varphi_1 + \varphi_2 \mright)} \sqrt{\frac{\gamma_3}{2}} \mright) \sm^b + \mleft( e^{i \varphi_1} \sqrt{\frac{\gamma_2}{2}} + \sqrt{\frac{\gamma_1}{2}} \mright) \sm^a, \\
H_{\rm L} &=& \frac{\sqrt{\gamma_1 \gamma_2}}{2} \sin \varphi_1 \frac{\sz^a}{2} + \frac{\sqrt{\gamma_3 \gamma_4}}{2} \sin \varphi_3 \frac{\sz^b}{2} \nn\\
&&+ \frac{1}{4i} \mleft[ \mleft( e^{i \mleft( \varphi_1 + \varphi_2 + \varphi_3 \mright)} \sqrt{\gamma_1 \gamma_4} + e^{i \mleft( \varphi_1 + \varphi_2 \mright)} \sqrt{\gamma_1 \gamma_3} + e^{i \mleft( \varphi_2 + \varphi_3 \mright)} \sqrt{\gamma_2 \gamma_4} + e^{i \varphi_2} \sqrt{\gamma_2 \gamma_3} \mright) \sp^a \sm^b - \text{H.c.} \mright].
\eea
The total triplet for the system is then $G_{\rm tot} = G_{\rm R} \boxplus G_{\rm L}$, with the components
\bea
\mathbf{S}_{\rm tot} &=& \begin{bmatrix} e^{i \mleft( \varphi_1 + \varphi_2 + \varphi_3 \mright)} & 0 \\ 0 & e^{i \mleft( \varphi_1 + \varphi_2 + \varphi_3 \mright)} \end{bmatrix}, \\
\mathbf{L}_{\rm tot} &=& \begin{bmatrix} \mleft( e^{i \mleft( \varphi_1 + \varphi_2 + \varphi_3 \mright)} \sqrt{\frac{\gamma_1}{2}} + e^{i \mleft( \varphi_2 + \varphi_3 \mright)} \sqrt{\frac{\gamma_2}{2}} \mright) \sm^a + \mleft( e^{i \varphi_3} \sqrt{\frac{\gamma_3}{2}} + \sqrt{\frac{\gamma_4}{2}} \mright) \sm^b \\ \mleft( \sqrt{\frac{\gamma_1}{2}} + e^{i \varphi_1} \sqrt{\frac{\gamma_2}{2}} \mright) \sm^a + \mleft( e^{i \mleft( \varphi_1 + \varphi_2 \mright)} \sqrt{\frac{\gamma_3}{2}} + e^{i \mleft( \varphi_1 + \varphi_2 + \varphi_3 \mright)} \sqrt{\frac{\gamma_4}{2}} \mright) \sm^b \end{bmatrix}, \\
H_{\rm tot} &=& \mleft( \omega_a + \sqrt{\gamma_1 \gamma_2} \sin \varphi_1 \mright) \frac{\sz^a}{2} + \mleft( \omega_b + \sqrt{\gamma_3 \gamma_4} \sin \varphi_3 \mright) \frac{\sz^b}{2} \nn\\
&&+ \frac{1}{2} \mleft[ \sqrt{\gamma_1 \gamma_4} \sin \mleft( \varphi_1 + \varphi_2 + \varphi_3 \mright) + \sqrt{\gamma_2 \gamma_4} \sin \mleft( \varphi_2 + \varphi_3 \mright) + \sqrt{\gamma_1 \gamma_3} \sin \mleft( \varphi_1 + \varphi_2 \mright) + \sqrt{\gamma_2 \gamma_3} \sin \varphi_2 \mright] \mleft(\sm^a \sp^b + \sp^a \sm^b \mright). \nn\\
\:
\eea

Continuing from the total triplet, we arrive at the master equation in the same way as in previous calculations:
\bea
\dot{\rho} &=& - i \comm{H_{\rm tot}}{\rho} + \lind{\mleft( e^{i \mleft( \varphi_1 + \varphi_2 + \varphi_3 \mright)} \sqrt{\frac{\gamma_1}{2}} + e^{i \mleft( \varphi_2 + \varphi_3 \mright)} \sqrt{\frac{\gamma_2}{2}} \mright) \sm^a + \mleft( e^{i \varphi_3} \sqrt{\frac{\gamma_3}{2}} + \sqrt{\frac{\gamma_4}{2}} \mright) \sm^b} \rho \nn\\
&&+ \lind{\mleft( e^{i \varphi_1} \sqrt{\frac{\gamma_2}{2}} + \sqrt{\frac{\gamma_1}{2}} \mright) \sm^a + \mleft( e^{i \mleft( \varphi_1 + \varphi_2 + \varphi_3 \mright)} \sqrt{\frac{\gamma_4}{2}} + e^{i \mleft( \varphi_1 + \varphi_2 \mright)} \sqrt{\frac{\gamma_3}{2}} \mright) \sm^b} \rho \nn\\
&=& - i \comm{H_{\rm tot}}{\rho} + \mleft(\gamma_1 + \gamma_2 + 2 \sqrt{\gamma_1 \gamma_2} \cos \varphi_1 \mright) \lind{\sm^a}\rho + \mleft( \gamma_3 + \gamma_4 + 2 \sqrt{\gamma_3 \gamma_4} \cos \varphi_3 \mright) \lind{\sm^b}\rho \nn\\
&&+ \mleft[ \sqrt{\gamma_1 \gamma_3} \cos \mleft( \varphi_1 + \varphi_2 \mright) + \sqrt{\gamma_1 \gamma_4} \cos \mleft( \varphi_1 + \varphi_2 + \varphi_3 \mright) + \sqrt{\gamma_2 \gamma_3} \cos \varphi_2 + \sqrt{\gamma_2 \gamma_4} \cos \mleft( \varphi_2 + \varphi_3 \mright) \mright] \nn\\
&&\times \mleft\{ \sm^a \rho \sp^b + \sm^b \rho \sp^a - \frac{1}{2}\mleft[ \mleft(\sp^a \sm^b + \sp^b \sm^a \mright) \rho + \rho \mleft( \sp^a \sm^b + \sp^b \sm^a \mright) \mright] \mright\}.
\label{eq:ME2SeparateGiantAtoms}
\eea
If we assume equal relaxation rates at all coupling points ($\gamma_1 = \gamma_2 = \gamma_3 = \gamma_4 \equiv \gamma$) and equal phases ($\varphi_1 = \varphi_2 = \varphi_3 \equiv \varphi$), \eqref{eq:ME2SeparateGiantAtoms} reduces to Eq.~(1) of the main text with the coefficients given in the fourth row of Table I.


\subsection{Braided giant atoms}

\begin{figure}
\centering
\includegraphics[width=\linewidth]{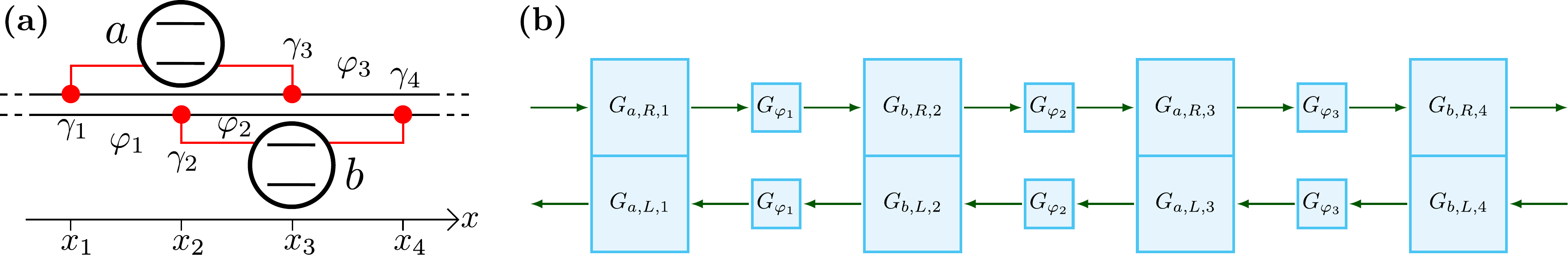}
\caption{Two braided giant atoms in an open waveguide. (a) A sketch showing the relevant parameters. (b) The input-output flows in the corresponding SLH calculation. \label{fig:2BraidedGiantAtomsSketchAndSLH}}
\end{figure}

We now consider two braided giant atoms. The setup, with definitions of all coupling strengths and phase shifts, is shown in \figref{fig:2BraidedGiantAtomsSketchAndSLH}(a). To calculate the SLH triplet for this setup, we follow the scheme sketched in \figref{fig:2BraidedGiantAtomsSketchAndSLH}(b). The triplets for the atoms are first decomposed into the following parts:
\bea
G_{a, \rm R, 1} &=& \mleft(1, \sqrt{\frac{\gamma_1}{2}}\sm^a, \omega_a \frac{\sz^a}{2} \mright), \\
G_{a, \rm R, 3} &=& \mleft(1, \sqrt{\frac{\gamma_3}{2}}\sm^a, 0 \mright), \\
G_{a, \rm L, 1} &=& \mleft(1, \sqrt{\frac{\gamma_1}{2}}\sm^a, 0 \mright), \\
G_{a, \rm L, 3} &=& \mleft(1, \sqrt{\frac{\gamma_3}{2}}\sm^a, 0 \mright), \\
G_{b, \rm R, 2} &=& \mleft(1, \sqrt{\frac{\gamma_2}{2}}\sm^b, \omega_b \frac{\sz^b}{2} \mright), \\
G_{b, \rm R, 4} &=& \mleft(1, \sqrt{\frac{\gamma_4}{2}}\sm^b, 0 \mright), \\
G_{b, \rm L, 2} &=& \mleft(1, \sqrt{\frac{\gamma_2}{2}}\sm^b, 0 \mright), \\
G_{b, \rm L, 4} &=& \mleft(1, \sqrt{\frac{\gamma_4}{2}}\sm^b, 0 \mright).
\eea
The triplet for the right-moving part is then given by
\be
G_{\rm R} = G_{b, \rm R, 4} \triangleleft G_{\varphi_3} \triangleleft G_{a, \rm R, 3} \triangleleft G_{\varphi_2} \triangleleft G_{b, \rm R, 2} \triangleleft G_{\varphi_1} \triangleleft G_{a, \rm R, 1}.
\ee
Comparing to the calculation for two small atoms in an open waveguide in \eqref{eq:GR2SmallAtoms}, we see that the first and last parts of this expression become
\bea
G_{b, \rm R, 4} \triangleleft G_{\varphi_3} \triangleleft G_{a, \rm R, 3} &=& \mleft( e^{i\varphi_3}, e^{i\varphi_3} \sqrt{\frac{\gamma_3}{2}}\sm^a + \sqrt{\frac{\gamma_4}{2}}\sm^b, \frac{\sqrt{\gamma_3 \gamma_4}}{4i} \mleft[ e^{i\varphi_3} \sm^a \sp^b - e^{-i\varphi_3} \sp^a \sm^b \mright] \mright), \\
G_{b, \rm R, 2} \triangleleft G_{\varphi_1} \triangleleft G_{a, \rm R, 1} &=& \mleft( e^{i\varphi_1}, e^{i\varphi_1} \sqrt{\frac{\gamma_1}{2}}\sm^a + \sqrt{\frac{\gamma_2}{2}}\sm^b, \omega_a \frac{\sz^a}{2} + \omega_b \frac{\sz^b}{2} + \frac{\sqrt{\gamma_1 \gamma_2}}{4i} \mleft[ e^{i\varphi_1} \sm^a \sp^b - e^{-i\varphi_1} \sp^a \sm^b \mright] \mright), \quad\quad
\eea
and further application of the series product rule leads to $G_{\rm R} = \mleft(\mathbf{S}_{\rm R}, \mathbf{L}_{\rm R}, H_{\rm R} \mright)$ with
\bea
\mathbf{S}_{\rm R} &=& e^{i \mleft( \varphi_1 + \varphi_2 + \varphi_3 \mright)}, \\
\mathbf{L}_{\rm R} &=& \mleft( e^{i \mleft( \varphi_1 + \varphi_2 + \varphi_3 \mright)} \sqrt{\frac{\gamma_1}{2}} + e^{i \varphi_3} \sqrt{\frac{\gamma_3}{2}} \mright) \sm^a + \mleft( e^{i \mleft(\varphi_2 + \varphi_3 \mright)} \sqrt{\frac{\gamma_2}{2}} + \sqrt{\frac{\gamma_4}{2}} \mright) \sm^b, \\
H_{\rm R} &=& \mleft[ \omega_a + \frac{\sqrt{\gamma_1 \gamma_3}}{2} \sin \mleft(\varphi_1 + \varphi_2 \mright) \mright] \frac{\sz^a}{2} + \mleft[ \omega_b + \frac{\sqrt{\gamma_2 \gamma_4}}{2} \sin \mleft( \varphi_2 + \varphi_3 \mright) \mright] \frac{\sz^b}{2} \nn\\
&&+ \frac{1}{4i} \mleft\{ \mleft[ \mleft( e^{i \varphi_1} \sqrt{\gamma_1 \gamma_2} + e^{i \mleft( \varphi_1 + \varphi_2 + \varphi_3 \mright)} \sqrt{\gamma_1 \gamma_4} + e^{i \varphi_3} \sqrt{\gamma_3 \gamma_4} \mright) \sm^a \sp^b + e^{i \varphi_2} \sqrt{\gamma_2 \gamma_3} \sp^a \sm^b \mright] - \text{H.c.} \mright\}.
\eea
From the symmetry that is apparent in \figref{fig:2BraidedGiantAtomsSketchAndSLH}(b), we can immediately deduce that the triplet for the left-moving part, $G_{\rm L} = \mleft(\mathbf{S}_{\rm L}, \mathbf{L}_{\rm L}, H_{\rm L} \mright)$, is given by removing the $\omega_j$ parts in the Hamiltonian, making the substitution $\varphi_1 \leftrightarrow \varphi_3$, and changing the non-$\varphi$ indices according to $a \leftrightarrow b$, $1 \leftrightarrow 4$ and $2 \leftrightarrow 3$ in the equations for $G_{\rm R}$, i.e.,
\bea
\mathbf{S}_{\rm L} &=& e^{i \mleft( \varphi_1 + \varphi_2 + \varphi_3 \mright)}, \\
\mathbf{L}_{\rm L} &=& \mleft( \sqrt{\frac{\gamma_1}{2}} + e^{i \mleft( \varphi_1 + \varphi_2 \mright)} \sqrt{\frac{\gamma_3}{2}} \mright) \sm^a + \mleft( e^{i \varphi_1} \sqrt{\frac{\gamma_2}{2}} + e^{i \mleft( \varphi_1 + \varphi_2 + \varphi_3 \mright)} \sqrt{\frac{\gamma_4}{2}} \mright) \sm^b, \\
H_{\rm L} &=& \mleft[ \frac{\sqrt{\gamma_1 \gamma_3}}{2} \sin \mleft(\varphi_1 + \varphi_2 \mright) \mright] \frac{\sz^a}{2} + \mleft[ \frac{\sqrt{\gamma_2 \gamma_4}}{2} \sin \mleft( \varphi_2 + \varphi_3 \mright) \mright] \frac{\sz^b}{2} \nn\\
&&+ \frac{1}{4i} \mleft\{ \mleft[ \mleft( e^{i \varphi_1} \sqrt{\gamma_1 \gamma_2} + e^{i \mleft( \varphi_1 + \varphi_2 + \varphi_3 \mright)} \sqrt{\gamma_1 \gamma_4} + e^{i \varphi_3} \sqrt{\gamma_3 \gamma_4} \mright) \sp^a \sm^b + e^{i \varphi_2} \sqrt{\gamma_2 \gamma_3} \sm^a \sp^b \mright] - \text{H.c.} \mright\}.
\eea
The total triplet for the system is then $G_{\rm tot} = G_{\rm R} \boxplus G_{\rm L}$, with the components
\bea
\mathbf{S}_{\rm tot} &=& \begin{bmatrix} e^{i \mleft( \varphi_1 + \varphi_2 + \varphi_3 \mright)} & 0 \\ 0 & e^{i \mleft( \varphi_1 + \varphi_2 + \varphi_3 \mright)} \end{bmatrix}, \\
\mathbf{L}_{\rm tot} &=& \begin{bmatrix} \mleft( e^{i \mleft( \varphi_1 + \varphi_2 + \varphi_3 \mright)} \sqrt{\frac{\gamma_1}{2}} + e^{i \varphi_3} \sqrt{\frac{\gamma_3}{2}} \mright) \sm^a + \mleft( e^{i \mleft( \varphi_2 + \varphi_3 \mright)} \sqrt{\frac{\gamma_2}{2}} + \sqrt{\frac{\gamma_4}{2}} \mright) \sm^b \\ \mleft( \sqrt{\frac{\gamma_1}{2}} + e^{i \mleft( \varphi_1 + \varphi_2 \mright)} \sqrt{\frac{\gamma_3}{2}} \mright) \sm^a + \mleft( e^{i \varphi_1} \sqrt{\frac{\gamma_2}{2}} + e^{i \mleft( \varphi_1 + \varphi_2 + \varphi_3 \mright)} \sqrt{\frac{\gamma_4}{2}} \mright) \sm^b \end{bmatrix}, \\
H_{\rm tot} &=& \mleft[ \omega_a + \sqrt{\gamma_1 \gamma_3} \sin \mleft( \varphi_1 + \varphi_2 \mright) \mright] \frac{\sz^a}{2} + \mleft[ \omega_b + \sqrt{\gamma_2 \gamma_4} \sin \mleft( \varphi_2 + \varphi_3 \mright) \mright] \frac{\sz^b}{2} \nn\\
&&+ \frac{1}{2} \mleft[ \sqrt{\gamma_1 \gamma_2} \sin \varphi_1 + \sqrt{\gamma_2 \gamma_3} \sin \varphi_2 + \sqrt{\gamma_3 \gamma_4} \sin \varphi_3 + \sqrt{\gamma_1 \gamma_4} \sin \mleft( \varphi_1 + \varphi_2 + \varphi_3 \mright) \mright] \mleft(\sm^a \sp^b + \sp^a \sm^b \mright).
\eea

Continuing from the total triplet, we arrive at the master equation in the same way as in previous calculations:
\bea
\dot{\rho} &=& - i \comm{H_{\rm tot}}{\rho} + \lind{\mleft( e^{i \mleft( \varphi_1 + \varphi_2 + \varphi_3 \mright)} \sqrt{\frac{\gamma_1}{2}} + e^{i \varphi_3} \sqrt{\frac{\gamma_3}{2}} \mright) \sm^a + \mleft( e^{i \mleft( \varphi_2 + \varphi_3 \mright)} \sqrt{\frac{\gamma_2}{2}} + \sqrt{\frac{\gamma_4}{2}} \mright) \sm^b} \rho \nn\\
&&+ \lind{\mleft( \sqrt{\frac{\gamma_1}{2}} + e^{i \mleft( \varphi_1 + \varphi_2 \mright)} \sqrt{\frac{\gamma_3}{2}} \mright) \sm^a + \mleft( e^{i \varphi_1} \sqrt{\frac{\gamma_2}{2}} + e^{i \mleft( \varphi_1 + \varphi_2 + \varphi_3 \mright)} \sqrt{\frac{\gamma_4}{2}} \mright) \sm^b} \rho \nn\\
&=& - i \comm{H_{\rm tot}}{\rho} + \mleft[ \gamma_1 + \gamma_3 + 2 \sqrt{\gamma_1 \gamma_3} \cos \mleft( \varphi_1 + \varphi_2 \mright) \mright] \lind{\sm^a}\rho + \mleft[ \gamma_2 + \gamma_4 + 2 \sqrt{\gamma_2 \gamma_4} \cos \mleft( \varphi_2 + \varphi_3 \mright) \mright] \lind{\sm^b}\rho \nn\\
&&+ \mleft[ \sqrt{\gamma_1 \gamma_2} \cos \varphi_1 + \sqrt{\gamma_2 \gamma_3} \cos \varphi_2 + \sqrt{\gamma_3 \gamma_4} \cos \varphi_3 + \sqrt{\gamma_1 \gamma_4} \cos \mleft( \varphi_1 + \varphi_2 + \varphi_3 \mright) \mright] \nn\\
&&\times \mleft\{ \sm^a \rho \sp^b + \sm^b \rho \sp^a - \frac{1}{2}\mleft[ \mleft(\sp^a \sm^b + \sp^b \sm^a \mright) \rho + \rho \mleft( \sp^a \sm^b + \sp^b \sm^a \mright) \mright] \mright\}.
\label{eq:ME2BraidedGiantAtoms}
\eea
If we assume equal relaxation rates at all coupling points ($\gamma_1 = \gamma_2 = \gamma_3 = \gamma_4 \equiv \gamma$) and equal phases ($\varphi_1 = \varphi_2 = \varphi_3 \equiv \varphi$), \eqref{eq:ME2BraidedGiantAtoms} reduces to Eq.~(1) of the main text with the coefficients given in the fifth row of Table I.


\subsection{Nested giant atoms}

\begin{figure}
\centering
\includegraphics[width=\linewidth]{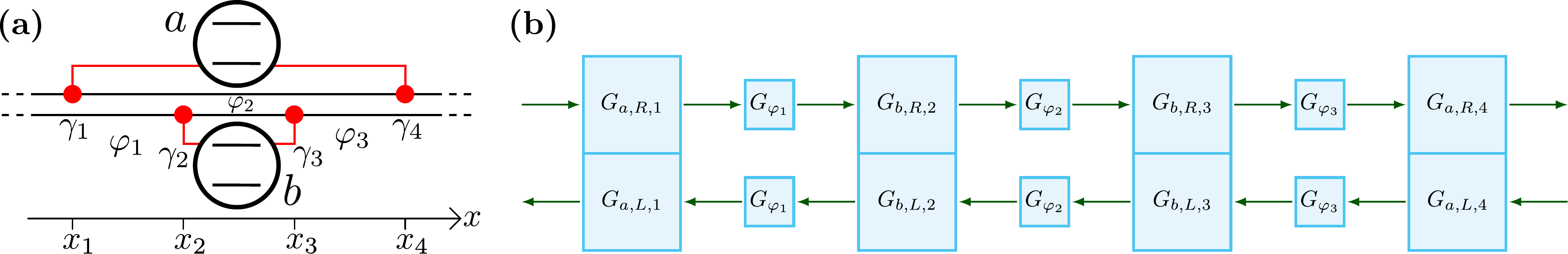}
\caption{Two nested giant atoms in an open waveguide. (a) A sketch showing the relevant parameters. (b) The input-output flows in the corresponding SLH calculation. \label{fig:2NestedGiantAtomsSketchAndSLH}}
\end{figure}

Finally, we consider two nested giant atoms. The setup, with definitions of all coupling strengths and phase shifts, is shown in \figref{fig:2NestedGiantAtomsSketchAndSLH}(a). To calculate the SLH triplet for this setup, we follow the scheme sketched in \figref{fig:2NestedGiantAtomsSketchAndSLH}(b). The triplets for the atoms are first decomposed into the following parts:
\bea
G_{a, \rm R, 1} &=& \mleft(1, \sqrt{\frac{\gamma_1}{2}}\sm^a, \omega_a \frac{\sz^a}{2} \mright), \\
G_{a, \rm R, 4} &=& \mleft(1, \sqrt{\frac{\gamma_4}{2}}\sm^a, 0 \mright), \\
G_{a, \rm L, 1} &=& \mleft(1, \sqrt{\frac{\gamma_1}{2}}\sm^a, 0 \mright), \\
G_{a, \rm L, 4} &=& \mleft(1, \sqrt{\frac{\gamma_4}{2}}\sm^a, 0 \mright), \\
G_{b, \rm R, 2} &=& \mleft(1, \sqrt{\frac{\gamma_2}{2}}\sm^b, \omega_b \frac{\sz^b}{2} \mright), \\
G_{b, \rm R, 3} &=& \mleft(1, \sqrt{\frac{\gamma_3}{2}}\sm^b, 0 \mright), \\
G_{b, \rm L, 2} &=& \mleft(1, \sqrt{\frac{\gamma_2}{2}}\sm^b, 0 \mright), \\
G_{b, \rm L, 3} &=& \mleft(1, \sqrt{\frac{\gamma_3}{2}}\sm^b, 0 \mright).
\eea
The triplet for the right-moving part is then given by
\be
G_{\rm R} = G_{a, \rm R, 4} \triangleleft G_{\varphi_3} \triangleleft G_{b, \rm R, 3} \triangleleft G_{\varphi_2} \triangleleft G_{b, \rm R, 2} \triangleleft G_{\varphi_1} \triangleleft G_{a, \rm R, 1}.
\ee
Repeated application of the series product rule, aided by comparison with results from previous calculations, leads to $G_{\rm R} = \mleft(\mathbf{S}_{\rm R}, \mathbf{L}_{\rm R}, H_{\rm R} \mright)$ with
\bea
\mathbf{S}_{\rm R} &=& e^{i \mleft( \varphi_1 + \varphi_2 + \varphi_3 \mright)}, \\
\mathbf{L}_{\rm R} &=& \mleft( e^{i \mleft( \varphi_1 + \varphi_2 + \varphi_3 \mright)} \sqrt{\frac{\gamma_1}{2}} + \sqrt{\frac{\gamma_4}{2}} \mright) \sm^a + \mleft( e^{i \mleft(\varphi_2 + \varphi_3 \mright)} \sqrt{\frac{\gamma_2}{2}} + e^{i \varphi_3} \sqrt{\frac{\gamma_3}{2}} \mright) \sm^b, \\
H_{\rm R} &=& \mleft[ \omega_a + \frac{\sqrt{\gamma_1 \gamma_4}}{2} \sin \mleft(\varphi_1 + \varphi_2 + \varphi_3 \mright) \mright] \frac{\sz^a}{2} + \mleft( \omega_b + \frac{\sqrt{\gamma_2 \gamma_3}}{2} \sin \varphi_2 \mright) \frac{\sz^b}{2} \nn\\
&&+ \frac{1}{4i} \mleft\{ \mleft[ \mleft( e^{i \varphi_1} \sqrt{\gamma_1 \gamma_2} + e^{i \mleft( \varphi_1 + \varphi_2 \mright)} \sqrt{\gamma_1 \gamma_3} \mright) \sm^a \sp^b + \mleft( e^{i \mleft( \varphi_2 + \varphi_3 \mright)} \sqrt{\gamma_2 \gamma_4} + e^{i \varphi_3} \sqrt{\gamma_3 \gamma_4} \mright) \sp^a \sm^b \mright] - \text{H.c.} \mright\}.
\eea
From the symmetry that is apparent in \figref{fig:2NestedGiantAtomsSketchAndSLH}(b), we can immediately deduce that the triplet for the left-moving part, $G_{\rm L} = \mleft(\mathbf{S}_{\rm L}, \mathbf{L}_{\rm L}, H_{\rm L} \mright)$, is given by removing the $\omega_j$ parts in the Hamiltonian, making the substitution $\varphi_1 \leftrightarrow \varphi_3$, and changing the non-$\varphi$ indices according to $1 \leftrightarrow 4$ and $2 \leftrightarrow 3$ in the equations for $G_{\rm R}$, i.e.,
\bea
\mathbf{S}_{\rm L} &=& e^{i \mleft( \varphi_1 + \varphi_2 + \varphi_3 \mright)}, \\
\mathbf{L}_{\rm L} &=& \mleft( \sqrt{\frac{\gamma_1}{2}} + e^{i \mleft( \varphi_1 + \varphi_2 + \varphi_3 \mright)} \sqrt{\frac{\gamma_4}{2}} \mright) \sm^a + \mleft( e^{i \varphi_1} \sqrt{\frac{\gamma_2}{2}} + e^{i \mleft(\varphi_1 + \varphi_2 \mright)} \sqrt{\frac{\gamma_3}{2}} \mright) \sm^b, \\
H_{\rm L} &=& \frac{\sqrt{\gamma_1 \gamma_4}}{2} \sin \mleft(\varphi_1 + \varphi_2 + \varphi_3 \mright) \frac{\sz^a}{2} + \frac{\sqrt{\gamma_2 \gamma_3}}{2} \sin \varphi_2 \frac{\sz^b}{2} \nn\\
&&+ \frac{1}{4i} \mleft\{ \mleft( e^{i \mleft( \varphi_2 + \varphi_3 \mright)} \sqrt{\gamma_2 \gamma_4} + e^{i \varphi_3} \sqrt{\gamma_3 \gamma_4} \mright) \sm^a \sp^b + \mleft[ \mleft( e^{i \varphi_1} \sqrt{\gamma_1 \gamma_2} + e^{i \mleft( \varphi_1 + \varphi_2 \mright)} \sqrt{\gamma_1 \gamma_3} \mright) \sp^a \sm^b \mright] - \text{H.c.} \mright\}.
\eea
The total triplet for the system is then $G_{\rm tot} = G_{\rm R} \boxplus G_{\rm L}$, with the components
\bea
\mathbf{S}_{\rm tot} &=& \begin{bmatrix} e^{i \mleft( \varphi_1 + \varphi_2 + \varphi_3 \mright)} & 0 \\ 0 & e^{i \mleft( \varphi_1 + \varphi_2 + \varphi_3 \mright)} \end{bmatrix}, \\
\mathbf{L}_{\rm tot} &=& \begin{bmatrix} \mleft( e^{i \mleft( \varphi_1 + \varphi_2 + \varphi_3 \mright)} \sqrt{\frac{\gamma_1}{2}} + \sqrt{\frac{\gamma_4}{2}} \mright) \sm^a + \mleft( e^{i \mleft(\varphi_2 + \varphi_3 \mright)} \sqrt{\frac{\gamma_2}{2}} + e^{i \varphi_3} \sqrt{\frac{\gamma_3}{2}} \mright) \sm^b \\ \mleft( \sqrt{\frac{\gamma_1}{2}} + e^{i \mleft( \varphi_1 + \varphi_2 + \varphi_3 \mright)} \sqrt{\frac{\gamma_4}{2}} \mright) \sm^a + \mleft( e^{i \varphi_1} \sqrt{\frac{\gamma_2}{2}} + e^{i \mleft(\varphi_1 + \varphi_2 \mright)} \sqrt{\frac{\gamma_3}{2}} \mright) \sm^b \end{bmatrix}, \\
H_{\rm tot} &=& \mleft[ \omega_a + \sqrt{\gamma_1 \gamma_4} \sin \mleft(\varphi_1 + \varphi_2 + \varphi_3 \mright) \mright] \frac{\sz^a}{2} + \mleft( \omega_b + \sqrt{\gamma_2 \gamma_3} \sin \varphi_2 \mright) \frac{\sz^b}{2} \nn\\
&&+ \frac{1}{2} \mleft[ \sqrt{\gamma_1 \gamma_2} \sin \varphi_1 + \sqrt{\gamma_1 \gamma_3} \sin \mleft( \varphi_1 + \varphi_2 \mright) + \sqrt{\gamma_2 \gamma_4} \sin \mleft( \varphi_2 + \varphi_3 \mright) + \sqrt{\gamma_3 \gamma_4} \sin \varphi_3 \mright] \mleft(\sm^a \sp^b + \sp^a \sm^b \mright).
\eea

Continuing from the total triplet, we arrive at the master equation in the same way as in previous calculations:
\bea
\dot{\rho} &=& - i \comm{H_{\rm tot}}{\rho} + \lind{\mleft( e^{i \mleft( \varphi_1 + \varphi_2 + \varphi_3 \mright)} \sqrt{\frac{\gamma_1}{2}} + \sqrt{\frac{\gamma_4}{2}} \mright) \sm^a + \mleft( e^{i \mleft(\varphi_2 + \varphi_3 \mright)} \sqrt{\frac{\gamma_2}{2}} + e^{i \varphi_3} \sqrt{\frac{\gamma_3}{2}} \mright) \sm^b} \rho \nn\\
&&+ \lind{\mleft( \sqrt{\frac{\gamma_1}{2}} + e^{i \mleft( \varphi_1 + \varphi_2 + \varphi_3 \mright)} \sqrt{\frac{\gamma_4}{2}} \mright) \sm^a + \mleft( e^{i \varphi_1} \sqrt{\frac{\gamma_2}{2}} + e^{i \mleft(\varphi_1 + \varphi_2 \mright)} \sqrt{\frac{\gamma_3}{2}} \mright) \sm^b} \rho \nn\\
&=& - i \comm{H_{\rm tot}}{\rho} + \mleft[ \gamma_1 + \gamma_4 + 2 \sqrt{\gamma_1 \gamma_4} \cos \mleft( \varphi_1 + \varphi_2 + \varphi_3 \mright) \mright] \lind{\sm^a}\rho + \mleft[ \gamma_2 + \gamma_3 + 2 \sqrt{\gamma_2 \gamma_3} \cos \varphi_2 \mright] \lind{\sm^b}\rho \nn\\
&&+ \mleft[ \sqrt{\gamma_1 \gamma_2} \cos \varphi_1 + \sqrt{\gamma_1 \gamma_3} \cos \mleft( \varphi_1 + \varphi_2 \mright) + \sqrt{\gamma_2 \gamma_4} \cos \mleft( \varphi_2 + \varphi_3 \mright) + \sqrt{\gamma_3 \gamma_4} \cos \varphi_3 \mright] \nn\\
&&\times \mleft\{ \sm^a \rho \sp^b + \sm^b \rho \sp^a - \frac{1}{2}\mleft[ \mleft(\sp^a \sm^b + \sp^b \sm^a \mright) \rho + \rho \mleft( \sp^a \sm^b + \sp^b \sm^a \mright) \mright] \mright\}.
\label{eq:ME2NestedGiantAtoms}
\eea
If we assume equal relaxation rates at all coupling points ($\gamma_1 = \gamma_2 = \gamma_3 = \gamma_4 \equiv \gamma$) and equal phases ($\varphi_1 = \varphi_2 = \varphi_3 \equiv \varphi$), \eqref{eq:ME2NestedGiantAtoms} reduces to Eq.~(1) of the main text with the coefficients given in the sixth row of Table I.


\section{Master equation for multiple giant atoms}
\label{sec:DerivationMostGeneralME}

We now turn to the general case of $N$ giant atoms, where each giant atom $j$ is coupled to the waveguide at $M_j$ points. The strength with which atom $j$ couples to the waveguide at connection point $j_n$ (coordinate $x_{j_n}$) is characterized by the relaxation rate $\gamma_{j_n}$. The positive phase acquired when moving from $j_n$ to $k_m$ is denoted $\varphi_{j_n, k_m}$. From the derivations for two giant atoms above, it is straightforward to generalize the SLH calculations to more atoms with more connection points. We first decompose the system into triplets for individual coupling points and phase shifts, interacting with right- and left-moving waves.

Looking at the right-moving part, 
\be
G_{\rm R} =  \mleft(\mathbf{S}_{\rm R}, \mathbf{L}_{\rm R}, H_{\rm R} \mright) = G_{N_{M_N}, \rm R} \triangleleft G_{\varphi_{\ldots, N_{M_N}}} \triangleleft \ldots \triangleleft G_{\varphi_{1_1, \ldots}} \triangleleft G_{1_1, \rm R},
\ee
we first note that the scattering matrix simply picks up a phase factor as we go from one connection point to the next. The total phase acquired from the first connection point, $1_1$, to the last connection point, $N_{M_N}$, is $\varphi_{1_1, N_{M_N}}$. For $\mathbf{L}_{\rm R}$, each connection point $j_n$ contributes $\sqrt{\gamma_{j_n}/2} \sm^{(j)}$, which then is multiplied by phase factors adding up to $\exp \mleft( i \varphi_{j_n, N_{M_N}} \mright)$ as we reach the last connection point. In $H_{\rm R}$, each pair of connection points $j_n$ and $j_m$ ($x_{j_n} < x_{j_m}$, i.e., $n < m$) belonging to the same atom $j$ contributes a term $\sqrt{\gamma_{j_n} \gamma_{j_m}} \sin \mleft(\varphi_{j_n, j_m} \mright) \sz^{(j)} / 4$, and each pair of connection points $j_n$ and $k_m$ ($x_{j_n} < x_{k_m}$) belonging to different atoms $j$ and $k$ contributes a term $\frac{1}{4i} \mleft[ \exp \mleft( i \varphi_{j_n, k_m} \mright) \sqrt{\gamma_{j_n} \gamma_{k_m}} \sm^{(j)} \sp^{(k)} - \text{H.c.} \mright]$. The result is 
\bea
\mathbf{S}_{\rm R} &=& \exp \mleft( i \varphi_{1_1, N_{M_N}} \mright), \\
\mathbf{L}_{\rm R} &=& \sum_{j=1}^N \sum_{n=1}^{M_j} \exp \mleft( i \varphi_{j_n, N_{M_N}} \mright) \sqrt{\frac{\gamma_{j_n}}{2}} \sm^{(j)}, \\
H_{\rm R} &=& \sum_{j=1}^N \mleft( \omega_j + \sum_{n = 1}^{M_j -1} \sum_{m = n +1}^{M_j} \frac{\sqrt{\gamma_{j_n} \gamma_{j_m}}}{2} \sin \varphi_{j_n, j_m} \mright) \frac{\sz^{(j)}}{2} \nn\\
&&+ \sum_{j = 1}^{N} \sum_{k \neq j} \sum_{n = 1}^{M_j} \sum_{\substack{m = 1 \\ x_{j_n} < x_{k_m}}}^{M_k} \frac{1}{4i} \sqrt{\gamma_{j_n} \gamma_{k_m}} \mleft[ \exp \mleft( i \varphi_{j_n, k_m} \mright) \sm^{(j)} \sp^{(k)} - \text{H.c.} \mright] .
\eea
From symmetry it follows that the components of the triplet $G_{\rm L}$ for the left-moving part are
\bea
\mathbf{S}_{\rm L} &=& \exp \mleft( i \varphi_{1_1, N_{M_N}} \mright), \\
\mathbf{L}_{\rm L} &=&  \sum_{j=1}^N \sum_{n=1}^{M_j} \exp \mleft( i \varphi_{1_1, j_n} \mright) \sqrt{\frac{\gamma_{j_n}}{2}} \sm^{(j)}, \\
H_{\rm L} &=& \sum_{j=1}^N \sum_{n = 1}^{M_j -1} \sum_{m = n +1}^{M_j} \frac{\sqrt{\gamma_{j_n} \gamma_{j_m}}}{2} \sin \varphi_{j_n, j_m} \frac{\sz^{(j)}}{2} \nn\\
&&+ \sum_{j = 1}^{N} \sum_{k \neq j} \sum_{n = 1}^{M_j} \sum_{\substack{m = 1 \\ x_{j_n} < x_{k_m}}}^{M_k} \frac{1}{4i} \sqrt{\gamma_{j_n} \gamma_{k_m}} \mleft[ \exp \mleft( i \varphi_{j_n, k_m} \mright) \sp^{(j)} \sm^{(k)} - \text{H.c.} \mright].
\eea

The total triplet for the system is then $G_{\rm tot} = G_{\rm R} \boxplus G_{\rm L}$, with the components
\bea
\mathbf{S}_{\rm tot} &=& \begin{bmatrix} \exp \mleft( i \varphi_{1_1, N_{M_N}} \mright) & 0 \\ 0 & \exp \mleft( i \varphi_{1_1,N_{M_N}} \mright) \end{bmatrix}, \label{eq:MostGeneralS} \\
\mathbf{L}_{\rm tot} &=& \begin{bmatrix} \sum_{j=1}^N \sum_{n=1}^{M_j} \exp \mleft( i \varphi_{j_n, N_{M_N}} \mright) \sqrt{\frac{\gamma_{j_n}}{2}} \sm^{(j)} \\ \sum_{j=1}^N \sum_{n=1}^{M_j} \exp \mleft( i \varphi_{1_1, j_n} \mright) \sqrt{\frac{\gamma_{j_n}}{2}} \sm^{(j)}  \end{bmatrix}, \\
H_{\rm tot} &=& \sum_{j=1}^N \mleft(\omega_j + \sum_{n = 1}^{M_j -1} \sum_{m = n +1}^{M_j} \sqrt{\gamma_{j_n} \gamma_{j_m}} \sin \varphi_{j_n, j_m} \mright) \frac{\sz^{(j)}}{2} \nn\\
&&+ \sum_{j = 1}^{N-1} \sum_{k = j + 1}^N \sum_{n = 1}^{M_j} \sum_{m = 1}^{M_k} \frac{\sqrt{\gamma_{j_n} \gamma_{k_m}}}{2} \sin \varphi_{j_n, k_m} \mleft( \sm^{(j)} \sp^{(k)} + \sp^{(j)} \sm^{(k)} \mright). \label{eq:MostGeneralH}
\eea
This leads to the master equation
\bea
\dot{\rho} &=& - i \comm{H_{\rm tot}}{\rho} + \lind{\sum_{j=1}^N \sum_{n=1}^{M_j} \exp \mleft( i \varphi_{j_n, N_{M_N}} \mright) \sqrt{\frac{\gamma_{j_n}}{2}} \sm^{(j)}} \rho + \lind{\sum_{j=1}^N \sum_{n=1}^{M_j} \exp \mleft( i \varphi_{1_1, j_n} \mright) \sqrt{\frac{\gamma_{j_n}}{2}} \sm^{(j)}} \rho \nn \\
&=& - i \comm{H_{\rm tot}}{\rho} + \sum_{j = 1}^N \sum_{n = 1}^{M_j} \sum_{m = 1}^{M_j} \frac{\sqrt{\gamma_{j_n} \gamma_{j_m}}}{2} \mleft[ \exp \mleft( i \varphi_{j_n, N_{M_N}} \mright) \exp \mleft( - i \varphi_{j_m, N_{M_N}} \mright) + \exp \mleft( i \varphi_{1_1, j_n} \mright) \exp \mleft( - i \varphi_{1_1, j_m} \mright) \mright] \lind{\sm^{(j)}}\rho \nn\\
&&+ \sum_{j = 1}^{N} \sum_{k \neq j} \sum_{n = 1}^{M_j} \sum_{m = 1}^{M_k} \frac{\sqrt{\gamma_{j_n} \gamma_{k_m}}}{2} \mleft[ \exp \mleft( i \varphi_{j_n, N_{M_N}} \mright) \exp \mleft( - i \varphi_{k_m, N_{M_N}} \mright) + \exp \mleft( i \varphi_{1_1, j_n} \mright) \exp \mleft( - i \varphi_{1_1, k_m} \mright) \mright] \nn\\
&& \times \mleft( \sm^{(j)} \rho \sp^{(k)} - \frac{1}{2}\mleft\{ \sp^{(j)} \sm^{(k)}, \rho \mright\} \mright) \nn\\
&=& - i \comm{H_{\rm tot}}{\rho} + \sum_{j = 1}^N \sum_{n = 1}^{M_j} \sum_{m = 1}^{M_j} \sqrt{\gamma_{j_n} \gamma_{j_m}} \cos \varphi_{j_n, j_m} \lind{\sm^{(j)}}\rho \nn\\
&&+ \sum_{j = 1}^{N-1} \sum_{k = j + 1}^N \sum_{n = 1}^{M_j} \sum_{m = 1}^{M_k} \sqrt{\gamma_{j_n} \gamma_{k_m}} \cos \varphi_{j_n, k_m} \mleft[ \mleft( \sm^{(j)} \rho \sp^{(k)} - \frac{1}{2}\mleft\{ \sp^{(j)} \sm^{(k)}, \rho \mright\} \mright) + \text{H.c.} \mright],
\eea
which is Eq.~(2) of the main text. Note that the individual and collective decay terms also could be written together in a more compact form:
%
\bea
&&\sum_{j = 1}^N \sum_{n = 1}^{M_j} \sum_{m = 1}^{M_j} \sqrt{\gamma_{j_n} \gamma_{j_m}} \cos \varphi_{j_n, j_m} \lind{\sm^{(j)}}\rho \nn\\
&+& \sum_{j = 1}^{N-1} \sum_{k = j + 1}^N \sum_{n = 1}^{M_j} \sum_{m = 1}^{M_k} \sqrt{\gamma_{j_n} \gamma_{k_m}} \cos \varphi_{j_n, k_m} \mleft[ \mleft( \sm^{(j)} \rho \sp^{(k)} - \frac{1}{2}\mleft\{ \sp^{(j)} \sm^{(k)}, \rho \mright\} \mright) + \text{H.c.} \mright] \nn\\
&=& \sum_{j = 1}^{N} \sum_{k = 1}^{N} \sum_{n = 1}^{M_j} \sum_{m = 1}^{M_k} \sqrt{\gamma_{j_n} \gamma_{k_m}} \cos \varphi_{j_n, k_m} \mleft( \sm^{(j)} \rho \sp^{(k)} - \frac{1}{2}\mleft\{ \sp^{(j)} \sm^{(k)}, \rho \mright\} \mright).
\eea
This should be compared to the master equation for $N$ \textit{small} atoms, each with individual relaxation rate $\gamma_j$, where the decay can be written as
\bea
&&\sum_{j, k = 1}^{N} \sqrt{\gamma_{j} \gamma_{k}} \cos \varphi_{j, k} \mleft( \sm^{(j)} \rho \sp^{(k)} - \frac{1}{2}\mleft\{ \sp^{(j)} \sm^{(k)}, \rho \mright\} \mright) \nn\\
&=& \sum_{j = 1}^{N} \gamma_j \lind{\sm^{(j)}}\rho + \sum_{j \neq k} \sqrt{\gamma_{j} \gamma_{k}} \cos \varphi_{j, k} \mleft( \sm^{(j)} \rho \sp^{(k)} - \frac{1}{2}\mleft\{ \sp^{(j)} \sm^{(k)}, \rho \mright\} \mright).
\eea
%


\section{Input-output relations}

For completeness, we note here that having access to the SLH components in Eqs.~(\ref{eq:MostGeneralS})-(\ref{eq:MostGeneralH}) makes it straightforward to derive the Hamiltonian and input-output relations in case a coherent drive or a Fock-state pulse is sent towards the giant atoms through the waveguide. For example, if a coherent drive at frequency $\omega_d$ with $\abssq{\alpha}$ photons per second is sent in from the left, the resulting triplet is
\be
G = \mleft(\mathbf{S}, \mathbf{L}, H \mright) = G_{\rm tot} \triangleleft \mleft( G_{\alpha} \boxplus \mathbbm{I}_1 \mright) = G_{\rm tot} \triangleleft \mleft( \begin{bmatrix} 1 & 0 \\ 0 & 1 \end{bmatrix}, \begin{bmatrix} \alpha \\ 0 \end{bmatrix}, 0 \mright),
\ee
which leads to
\bea
\mathbf{S} &=& \begin{bmatrix} \exp \mleft( i \varphi_{1_1,N_{M_N}} \mright) & 0 \\ 0 & \exp \mleft( i \varphi_{1_1,N_{M_N}} \mright) \end{bmatrix}, \\
\mathbf{L} &=& \begin{bmatrix} \alpha \exp \mleft( i \varphi_{1_1,N_{M_N}} \mright) + \sum_{j=1}^N \sum_{n=1}^{M_j} \exp \mleft( i \varphi_{j_n, N_{M_N}} \mright) \sqrt{\frac{\gamma_{j_n}}{2}} \sm^{(j)} \\ \sum_{j=1}^N \sum_{n=1}^{M_j} \exp \mleft( i \varphi_{1_1, j_n} \mright) \sqrt{\frac{\gamma_{j_n}}{2}} \sm^{(j)}  \end{bmatrix}, \label{eq:MostGeneralDrivenL} \\
H &=& \sum_{j=1}^N \mleft( \Delta_j + \sum_{n = 1}^{M_j -1} \sum_{m = n +1}^{M_j} \sqrt{\gamma_{j_n} \gamma_{j_m}} \sin \varphi_{j_n, j_m} \mright) \frac{\sz^{(j)}}{2} \nn\\
&&+ \sum_{j = 1}^{N-1} \sum_{k = j + 1}^N \sum_{n = 1}^{M_j} \sum_{m = 1}^{M_k} \frac{\sqrt{\gamma_{j_n} \gamma_{k_m}}}{2} \sin \varphi_{j_n, k_m} \mleft( \sm^{(j)} \sp^{(k)} + \sp^{(j)} \sm^{(k)} \mright) \nn \\
&&+ \frac{1}{2i} \mleft[ \alpha \exp \mleft( i \varphi_{1_1,N_{M_N}} \mright) \sum_{j=1}^N \sum_{n=1}^{M_j} \exp \mleft( - i \varphi_{j_n, N_{M_N}} \mright) \sqrt{\frac{\gamma_{j_n}}{2}} \sp^{(j)} - \text{H.c.} \mright] ,
\eea
where we have moved to a frame rotating with $\omega_d$, introducing the notation $\Delta_j = \omega_j - \omega_d$. The master equation for the driven system then becomes
\bea
\dot{\rho} &=& - i \comm{H_{\rm driven}}{\rho} + \sum_{j = 1}^N \sum_{n = 1}^{M_j} \sum_{m = 1}^{M_j} \sqrt{\gamma_{j_n} \gamma_{j_m}} \cos \varphi_{j_n, j_m} \lind{\sm^{(j)}}\rho \nn\\
&&+ \sum_{j = 1}^{N-1} \sum_{k = j + 1}^{N} \sum_{n = 1}^{M_j} \sum_{m = 1}^{M_k} \sqrt{\gamma_{j_n} \gamma_{k_m}} \cos \varphi_{j_n, k_m} \mleft[ \mleft( \sm^{(j)} \rho \sp^{(k)} - \frac{1}{2}\mleft\{ \sp^{(j)} \sm^{(k)}, \rho \mright\} \mright) + \text{H.c.} \mright],
\eea
where
\bea
H_{\rm driven} &=& \sum_{j=1}^N \mleft( \Delta_j + \sum_{n = 1}^{M_j -1} \sum_{m = n +1}^{M_j} \sqrt{\gamma_{j_n} \gamma_{j_m}} \sin \varphi_{j_n, j_m} \mright) \frac{\sz^{(j)}}{2} \nn\\
&&+ \sum_{j = 1}^{N-1} \sum_{k = j + 1}^N \sum_{n = 1}^{M_j} \sum_{m = 1}^{M_k} \frac{\sqrt{\gamma_{j_n} \gamma_{k_m}}}{2} \sin \varphi_{j_n, k_m} \mleft( \sm^{(j)} \sp^{(k)} + \sp^{(j)} \sm^{(k)} \mright) \nn \\
&& -i \sum_{j=1}^N \sum_{n=1}^{M_j} \sqrt{\frac{\gamma_{j_n}}{2}} \mleft[ \alpha \exp \mleft( i \varphi_{1_1, j_n} \mright) \sp^{(j)} - \alpha^* \exp \mleft( - i \varphi_{1_1, j_n} \mright) \sm^{(j)} \mright].
\eea
The output traveling to the right from the atoms is given by the upper row of $\mathbf{L}$ in \eqref{eq:MostGeneralDrivenL} and the output traveling to the left is given by the lower row of $\mathbf{L}$ in the same equation. From this, we can obtain the amplitude transmission coefficient $t$ and the amplitude reflection coefficient $r$:
\bea
t &=& \frac{\expec{L_1}}{\alpha} = \exp \mleft( i \varphi_{1_1,N_{M_N}} \mright) + \frac{1}{\alpha} \sum_{j=1}^N \sum_{n=1}^{M_j} \exp \mleft( i \varphi_{j_n, N_{M_N}} \mright) \sqrt{\frac{\gamma_{j_n}}{2}} \expec{\sm^{(j)}}, \\
r &=& \frac{\expec{L_2}}{\alpha} = \frac{1}{\alpha} \sum_{j=1}^N \sum_{n=1}^{M_j} \exp \mleft( i \varphi_{1_1, j_n} \mright) \sqrt{\frac{\gamma_{j_n}}{2}} \expec{\sm^{(j)}}.
\eea
%


\section{Connection between exchange interaction, individual decay, and collective decay for giant atoms}

In this section, we give detailed proofs for the statements in the main text regarding the connections between the exchange interaction $g$, the individual decays $\Gamma_j$, and the collective decays $\Gamma_{{\rm coll}, j, k}$. We first consider two giant atoms with two connection points each and then generalize to multiple atoms with multiple connections points.


\subsection{Two atoms, two connection points}

For two atoms with two connection points, there are three distinct geometries, shown in Fig.~1(c)-(e).
Before treating these cases one by one, we note that the individual decay rates in all these setups can be written
\be
\Gamma_j = \gamma_{j,1} + \gamma_{j,2} + 2 \sqrt{\gamma_{j,1} \gamma_{j,2}} \cos \varphi_j,
\ee
where $\gamma_{j,1}$ and $\gamma_{j,2}$ are the relaxation rates characterizing the coupling of atom $j$ to the waveguide at connection points 1 and 2, respectively, and $\varphi_j$ is the phase acquired traveling between these two connection points. From the inequality between the arithmetic and geometric means,
\be
\frac{\gamma_{j,1} + \gamma_{j,2}}{2} \geq \sqrt{\gamma_{j,1} \gamma_{j,2}},
\ee
we see that $\Gamma_j$ can only become zero if $\gamma_{j,1} = \gamma_{j,2}$. Since we are interested in exactly those cases where the decay rates for different setups go to zero, we will use $\gamma_{j,1} = \gamma_{j,2} \equiv \gamma_j$ throughout this subsection.


\subsubsection{Separate giant atoms}

For separate giant atoms, the master equation is given in \eqref{eq:ME2SeparateGiantAtoms}, from which we read off
\bea
g &\propto& \sqrt{\gamma_a \gamma_b} \mleft[ \sin \mleft( \varphi_1 + \varphi_2 \mright) + \sin \mleft( \varphi_1 + \varphi_2 + \varphi_3 \mright) + \sin \varphi_2 + \sin \mleft( \varphi_2 + \varphi_3 \mright) \mright] \\
\Gamma_a &\propto& \gamma_a (1+\cos \varphi_1), \\
\Gamma_b &\propto& \gamma_b (1+\cos \varphi_3), \\
\Gamma_{\rm coll} &\propto& \sqrt{\gamma_a \gamma_b} \mleft[ \cos \mleft( \varphi_1 + \varphi_2 \mright) + \cos \mleft( \varphi_1 + \varphi_2 + \varphi_3 \mright) + \cos \varphi_2 + \cos \mleft( \varphi_2 + \varphi_3 \mright) \mright].
\eea
If we set the individual decay terms to zero, this implies $\varphi_1 = (2n + 1) \pi$ and $\varphi_3 = (2m + 1) \pi$ with $n, m \in \Z$. Inserting this into the expression for the collective decay gives
\bea
\Gamma_{\rm coll} &\propto& \mleft\{ \cos \mleft[ (2n + 1) \pi + \varphi_2 \mright] + \cos \mleft[ (2n + 1) \pi + \varphi_2 + (2m + 1) \pi \mright] + \cos \varphi_2 + \cos \mleft[ \varphi_2 + (2m + 1) \pi \mright] \mright\} \nn\\
&=& \cos \mleft( \varphi_2 + \pi \mright) + \cos \varphi_2 + \cos \varphi_2 + \cos \mleft (\varphi_2 + \pi \mright) = 2 \mleft[ \cos \varphi_2 + \cos \mleft( \varphi_2 + \pi \mright) \mright] = 2 \mleft( \cos\varphi_2 - \cos \varphi_2 \mright) = 0.
\eea
For the exchange interaction, we similarly obtain
\bea
g &\propto& \mleft\{ \sin \mleft[ (2n + 1) \pi + \varphi_2 \mright] + \sin \mleft[ (2n + 1) \pi + \varphi_2 + (2m + 1) \pi \mright] + \sin \varphi_2 + \sin \mleft[ \varphi_2 + (2m + 1) \pi \mright] \mright\} \nn\\
&=& \sin \mleft( \varphi_2 + \pi \mright) + \sin \varphi_2 + \sin \varphi_2 + \sin \mleft(\varphi_2 + \pi \mright) = 2 \mleft[ \sin\varphi_2 + \sin \mleft(\varphi_2 + \pi \mright) \mright] = 2 \mleft( \sin \varphi_2 - \sin \varphi_2 \mright) = 0.
\eea
Thus we conclude that if the individual decays are set to zero for separate giant atoms, \textit{both} the collective decay and the exchange interaction will also be zero.

If we instead set the exchange interaction to zero, neither the individual nor the collective decays need become zero. A counter-example is provided by making all $\varphi_j$ integer multiples of $2\pi$; this maximizes the individual and collective decays while the exchange interaction becomes zero.

If we set the collective decay to zero, neither the individual decays nor the exchange interaction need become zero. A counter-example is $\varphi_1 = \varphi_3 = 2\pi$ and $\varphi_2 = \pi/2$, which maximizes both the individual decays and the exchange interaction while making the collective decay zero.


\subsubsection{Braided giant atoms}

For separate giant atoms, the master equation is given in \eqref{eq:ME2BraidedGiantAtoms}, from which we read off
\bea
g &\propto& \sqrt{\gamma_a \gamma_b} \mleft[ \sin \varphi_1 + \sin \varphi_2 + \sin \varphi_3 + \sin \mleft( \varphi_1 + \varphi_2 + \varphi_3 \mright) \mright] \\
\Gamma_a &\propto& \gamma_a \mleft[ 1+\cos \mleft( \varphi_1 + \varphi_2 \mright) \mright], \\
\Gamma_b &\propto& \gamma_b \mleft[ 1+\cos \mleft( \varphi_2 + \varphi_3 \mright) \mright], \\
\Gamma_{\rm coll} &\propto& \sqrt{\gamma_a \gamma_b} \mleft[ \cos \varphi_1 + \cos \varphi_2 + \cos \varphi_3 + \cos \mleft( \varphi_1 + \varphi_2 + \varphi_3 \mright) \mright].
\eea
If we set the individual decay terms to zero, this implies $\varphi_1 + \varphi_2 = (2n + 1) \pi$ and $\varphi_2 + \varphi_3 = (2m + 1) \pi$ with $n, m \in \Z$. Solving in terms of $\varphi_2$, we obtain $\varphi_1 = (2n + 1) \pi - \varphi_2$ and $\varphi_3 = (2m + 1) \pi - \varphi_2$, which inserted into the collective decay leads to
\be
\Gamma_{\rm coll} \propto \mleft[\cos \mleft( \pi - \varphi_2 \mright) + \cos \mleft( - \varphi_2 \mright) + \cos \varphi_2 + \cos \mleft( \pi - \varphi_2 \mright) \mright] = 0.
\ee
For the exchange interaction, we similarly obtain
\be
g \propto \mleft[ \sin \mleft( \pi - \varphi_2 \mright) + \sin \mleft( - \varphi_2 \mright) + \sin \varphi_2 + \sin \mleft( \pi - \varphi_2 \mright) \mright] = 2 \sin \varphi_2.
\label{eq:gBraidedAtoms}
\ee
This is possibly \textit{the most important result} of the present work. Since $\varphi_2$ is a free parameter (we only fix $\varphi_1$ and $\varphi_3$ to make the individual decays zero), we can engineer a \textit{large} exchange interaction, and \textit{choose its sign}, even though both the individual decays and the collective decay are zero.

If we instead set the exchange interaction to zero for the setup with braided giant atoms, neither the individual nor the collective decays need become zero. The same counter-example as for separate giant atoms works here as well: making all $\varphi_j$ integer multiples of $2\pi$ maximizes the individual and collective decays while the exchange interaction becomes zero.

If we set the collective decay to zero, this does \textit{not} mean that the exchange interaction or all of the individual decay will be zero. A counter-example is $\varphi_1 = 2\pi/3$, $\varphi_2 = \pi/2$, and $\varphi_3 = \pi/3$ which makes the collective decay zero, but the exchange interaction and the individual decays all become nonzero.


\subsubsection{Nested giant atoms}

For nested giant atoms, the master equation is given in \eqref{eq:ME2NestedGiantAtoms}, from which we read off
\bea
g &\propto& \sqrt{\gamma_a \gamma_b} \mleft[ \sin \varphi_1 + \sin \mleft( \varphi_1 + \varphi_2 \mright) + \sin \mleft( \varphi_2 + \varphi_3 \mright) + \sin \varphi_3 \mright] \\
\Gamma_a &\propto& \gamma_a \mleft[ 1+\cos \mleft( \varphi_1 + \varphi_2 + \varphi_3 \mright) \mright], \\
\Gamma_b &\propto& \gamma_b \mleft( 1+\cos \varphi_2 \mright), \\
\Gamma_{\rm coll} &\propto& \sqrt{\gamma_a \gamma_b} \mleft[ \cos \varphi_1 + \cos \mleft( \varphi_1 + \varphi_2 \mright) + \cos \mleft( \varphi_2 + \varphi_3 \mright) + \cos \varphi_3 \mright].
\eea
If we set the individual decay terms to zero, this implies $\varphi_2 = (2n + 1) \pi$ and $\varphi_1 + \varphi_2 + \varphi_3 = (2m + 1) \pi$ with $n, m \in \Z$. Using these two constraints to express $\varphi_3$ in terms of $\varphi_1$, the collective decay becomes
\be
\Gamma_{\rm coll} \propto \mleft[ \cos \varphi_1 + \cos \mleft(\varphi_1 + \pi \mright) + \cos \mleft(\pi - \varphi_1 \mright) + \cos \mleft( - \varphi_1 \mright) \mright] = 0.
\ee
For the exchange interaction, we similarly obtain
\be
g \propto \mleft[ \sin \varphi_1 + \sin \mleft( \varphi_1 + \pi \mright) + \sin \mleft( \pi - \varphi_1 \mright) + \sin \mleft( - \varphi_1 \mright) \mright] = 0.
\ee
Thus we conclude that if the individual decays are set to zero for nested giant atoms, \textit{both} the collective decay and the exchange interaction will also be zero.

If we instead set the exchange interaction to zero for the setup with nested giant atoms, neither the individual nor the collective decays need become zero. The same counter-example as for separate and braided giant atoms works here as well: making all $\varphi_j$ integer multiples of $2\pi$ maximizes the individual and collective decays while the exchange interaction becomes zero.

If we set the collective decay to zero, this does not mean that the exchange interaction, nor all of the individual decay terms, will be zero. A simple counter-example is $\varphi_1 = \varphi_3 = \pi/2$ and $\varphi_2 = 2\pi$, which makes the collective decay zero, maximizes the exchange interaction, maximizes the individual decay of atom $b$ (the inner atom) and makes the individual decay of atom $a$ zero.


\subsection{Multiple atoms, two connection points}

From Eq.~(2) in the main text, we know that all interactions occur between pairs of atoms.
If we have multiple atoms with two connection points each, we can consider two atoms at a time and check whether they are separate, braided or nested. When this classification has been determined, the conclusions from the previous subsection applies to the master-equation terms for this pair of atoms.

For the case of two braided atoms in the previous subsection, it was clear that $g$ could be \textit{set arbitrarily} through the free parameter $\varphi_2$ while using $\varphi_1$ and $\varphi_3$ to ensure that all the individual decay terms, and thus also the collective decay, remained zero. For more than two braided atoms, setting all decay terms to zero may introduce too many constraints to also allow for arbitrary control of all $g_{j,k}$. Below, we explicitly show whether this is the case for the two setups in Figs.~3 and 4 of the main text, i.e., for atoms connected in a 1D chain with nearest-neighbor couplings or in a setup allowing pairwise couplings between all atoms.


\subsubsection{1D chain with nearest-neighbor couplings}

\begin{figure}
\centering
\includegraphics[width=\linewidth]{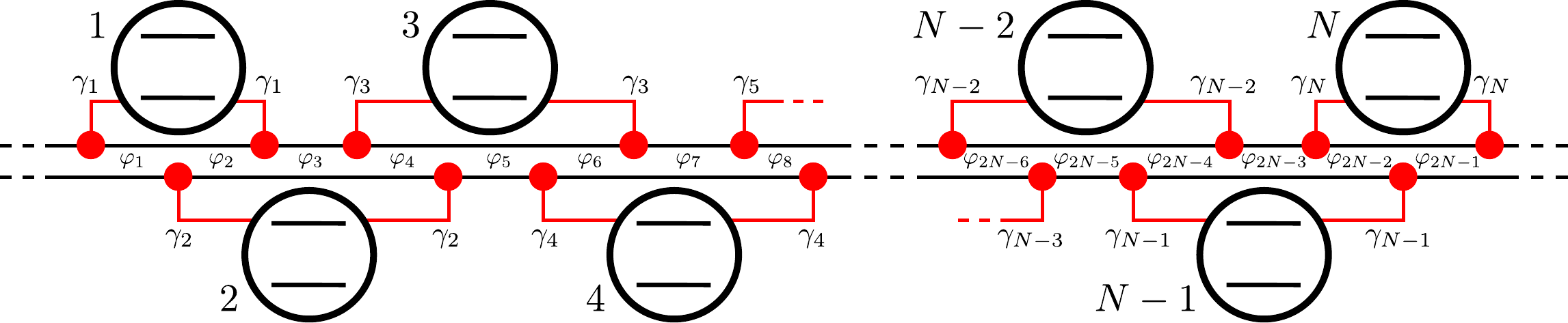}
\caption{A sketch showing the relevant parameters for a 1D chain of giant atoms with nearest-neighbor couplings protected from relaxation.
\label{fig:1DNearestNeighbourWithNotation}}
\end{figure}

The setup for a 1D chain of $N$ braided giant atoms is shown in \figref{fig:1DNearestNeighbourWithNotation} with all phases and coupling strengths marked. Just as before, we note that the individual decay rate $\Gamma_j$ for the $j$th atom only can become zero if the couples to the waveguide with equal strengths, denoted $\gamma_j$, at each of its two connection points. The $N$ constraints that make all individual decay rates zero are then
\bea
\varphi_1 + \varphi_2 &=& (2n_1 + 1) \pi, \label{eq:1DConstraint1} \\
\varphi_2 + \varphi_3 + \varphi_4 &=& (2n_2 + 1) \pi, \\
\varphi_4 + \varphi_5 + \varphi_6 &=& (2n_3 + 1) \pi, \\
&\vdots& \nn \\
\varphi_{2 N - 6} + \varphi_{2 N - 5} + \varphi_{2 N - 4} &=& (2n_{N-2} + 1) \pi, \\
\varphi_{2 N - 4} + \varphi_{2 N - 3} + \varphi_{2 N - 2} &=& (2n_{N-1} + 1) \pi, \\
\varphi_{2 N - 2} + \varphi_{2 N - 1} &=& (2n_N + 1) \pi, \label{eq:1DConstraintN}
\eea
where $n_1, n_2, \ldots, n_N \in \Z$. From the calculation for two braided giant atoms in \eqref{eq:gBraidedAtoms}, we see that the nearest-neighbor couplings can be expressed as
\be
g_{j, j+1} = \sqrt{\gamma_j \gamma_{j+1}} \sin \varphi_{2j}.
\ee
This implies that all the nearest-neighbor couplings can be tuned individually by choosing $\varphi_{2j}$ for $j = 1, 2, \ldots, N-1$. Since there are in total $2N-1$ phases, this leaves $N$ free parameters, $\varphi_{2j-1}$ for $j = 1, 2, \ldots, N$, which is exactly what is needed to satisfy the constraints in Eqs.~(\ref{eq:1DConstraint1})-(\ref{eq:1DConstraintN}) and make all the individual decay rates zero.


\subsubsection{Pairwise coupling between all atoms}

\begin{figure}
\centering
\includegraphics[width=0.45\linewidth]{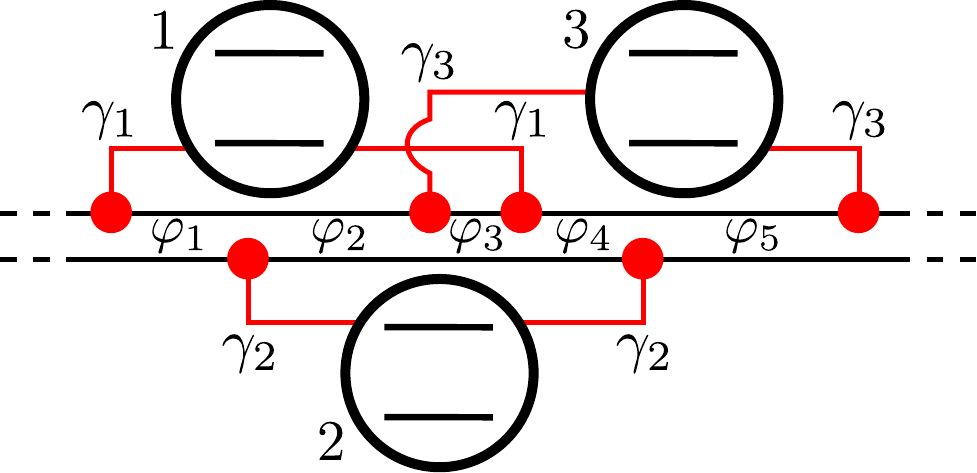}
\caption{A sketch showing the relevant parameters for three giant atoms braided such that all possible pairs of atoms can achieve decoherence-free exchange interaction.
\label{fig:3GiantAtomsAllCoupledWithNotation}}
\end{figure}

If we braid more giant atoms (with two connection points each) than just the nearest neighbors in a 1D chain, as shown for three giant atoms in Fig.~4 in the main text and again with more detailed notation in \figref{fig:3GiantAtomsAllCoupledWithNotation}, the number of constraints that need to be satisfied to ensure that all decay rates are zero remains unchanged ($N$ constraints for $N$ atoms).
The number of connection points is still $2N$, so the number of phases that can be set is still $2 N - 1$. However, if all atoms are to be pairwise connected, the number of exchange couplings becomes $N(N-1)/2$, which, for $N \geq 3$, exceeds the number of free parameters, $N - 1$, remaining after satisfying the constraints for zero relaxation. Thus, we will not be able to set all exchange-coupling strengths arbitrarily when the connectivity exceeds that of the 1D chain treated in the previous subsection.

For completeness, we here show explicitly some coupling-strength configurations that can be achieved for the setup with three pairwise connected atoms, shown in \figref{fig:3GiantAtomsAllCoupledWithNotation}. The constraints ensuring zero relaxation through the waveguide are
\bea
\varphi_1 + \varphi_2 + \varphi_3 &=& (2n_1 + 1) \pi, \label{eq:3CoupledAtomsConstraint1} \\
\varphi_2 + \varphi_3 + \varphi_4 &=& (2n_2 + 1) \pi, \label{eq:3CoupledAtomsConstraint2} \\
\varphi_3 + \varphi_4 + \varphi_5 &=& (2n_3 + 1) \pi, \label{eq:3CoupledAtomsConstraint3}
\eea
where $n_1, n_2, n_3 \in \Z$. The pairwise couplings are
\bea
g_{1,2} &=& \sqrt{\gamma_1 \gamma_2} \sin \mleft( \varphi_2 + \varphi_3 \mright), \\
g_{1,3} &=& \sqrt{\gamma_1 \gamma_3} \sin \varphi_3, \\
g_{2,3} &=& \sqrt{\gamma_2 \gamma_3} \sin \mleft( \varphi_3 + \varphi_4 \mright).
\eea
The constraints in Eqs.~(\ref{eq:3CoupledAtomsConstraint1}) and (\ref{eq:3CoupledAtomsConstraint3}) can be satisfied without affecting the exchange interactions by choosing suitable $\varphi_1$ and $\varphi_5$, respectively. If $\gamma_1 = \gamma_2 = \gamma_3$, we can make all pairwise couplings equal and satisfy the constraint in \eqref{eq:3CoupledAtomsConstraint2} at the same time by choosing $\varphi_2 = \varphi_3 = \varphi_4 = \pi / 3$. We can also satisfy the constraint in \eqref{eq:3CoupledAtomsConstraint2} and engineer the couplings to obey $g_{1,2} = g_{2,3} = - g_{1,3}$ by choosing $\varphi_2 = \varphi_4 = 4 \pi / 3$ and $\varphi_3 = \pi / 3$.


\subsection{Multiple atoms, multiple connection points}

For the general case, with $N$ giant atoms, where atom $j$ has $M_j$ connection points, we can still divide each pair of atoms $j$ and $k$ into one of the three categories separate, braided, and nested. Explicitly, the atoms are separate if $x_{j_{M_j}} < x_{k_1}$, i.e., if all connection points of atom $j$ are situated to the left of all connection points of atom $k$. The atoms are nested if, for some $n$, $x_{j_n} < x_{k_m} < x_{j_{n+1}}$ $\forall m$, i.e., if all connection points of atom $k$ are situated in-between two subsequent connection points of atom $j$. All other cases are braided atoms, where some of the connection points of atom $j$ are situated in-between some of the connection points of atom $k$.

To prove the connections between exchange interaction, individual decay, and collective decay for these general cases, it is convenient to look back at the derivation of the general master equation in \secref{sec:DerivationMostGeneralME}. We observe that the individual decay rate $\Gamma_j$ of atom $j$ can be expressed as~\cite{Kockum2014}
\be
\Gamma_j = \abssq{\sum_{n = 1}^{M_j} \sqrt{\gamma_{j_n}} \exp \mleft( i \varphi_{j_1, j_n}\mright)} = \abssq{\sum_{n = 1}^{M_j} \sqrt{\gamma_{j_n}} \exp \mleft( i \varphi_{j_n, j_{M_j}}\mright)},
\ee
since
\be
\sum_{n = 1}^{M_j} \sqrt{\gamma_{j_n}} \exp \mleft( i \varphi_{j_1, j_n} \mright) = \mleft\{ \exp \mleft( - i \varphi_{j_1, j_{M_j}} \mright) \sum_{n = 1}^{M_j} \sqrt{\gamma_{j_n}} \exp \mleft( i \varphi_{j_n, j_{M_j}} \mright) \mright\}^*.
\ee
This means that $\Gamma_j = 0$ implies $\sum_{n = 1}^{M_j} \sqrt{\gamma_{j_n}} \exp \mleft( i \varphi_{j_1, j_n}\mright) = 0$ and $\sum_{n = 1}^{M_j} \sqrt{\gamma_{j_n}} \exp \mleft( i \varphi_{j_n, j_{M_j}} \mright) = 0$.

From the derivation in \secref{sec:DerivationMostGeneralME}, we see that the collective coupling can be written as
\bea
\Gamma_{{\rm coll}, j, k} &=& \sum_{n = 1}^{M_j} \sum_{m = 1}^{M_k} \frac{\sqrt{\gamma_{j_n} \gamma_{k_m}}}{2} \mleft[ \exp \mleft( i \varphi_{j_n, N_{M_N}} \mright) \exp \mleft( - i \varphi_{k_m, N_{M_N}} \mright) + \exp \mleft( i \varphi_{1_1, j_n} \mright) \exp \mleft( - i \varphi_{1_1, k_m} \mright) \mright] \nn\\
&=& \frac{1}{2}  \exp \mleft( i \varphi_{j_{M_j}, N_{M_N}} \mright) \underbrace{\sum_{n = 1}^{M_j} \sqrt{\gamma_{j_n}} \exp \mleft( i \varphi_{j_n, j_{M_j}} \mright)}_{= 0 \:\text{if}\: \Gamma_j = 0} \sum_{m = 1}^{M_k} \sqrt{\gamma_{k_m}} \exp \mleft( - i \varphi_{k_m, N_{M_N}} \mright) \nn\\
&&+ \frac{1}{2}  \exp \mleft( i \varphi_{1_1, j_1} \mright) \underbrace{\sum_{n = 1}^{M_j} \sqrt{\gamma_{j_n}} \exp \mleft( i \varphi_{j_1, j_n} \mright)}_{= 0 \:\text{if}\: \Gamma_j = 0} \sum_{m = 1}^{M_k} \sqrt{\gamma_{k_m}} \exp \mleft( - i \varphi_{1_1, k_m} \mright),
\eea
and hence we conclude that $\Gamma_j = 0$ implies $\Gamma_{{\rm coll}, j, k} = 0 \:\forall k$.

Finally, we study the exchange interaction $g_{j,k}$ for a pair of atoms where, without loss of generality, we assume $x_{j_1} < x_{k_1}$. Using similar manipulations as above, it can be written
\bea
g_{j,k} &=& \sum_{n = 1}^{M_j} \sum_{m = 1}^{M_k} \frac{\sqrt{\gamma_{j_n} \gamma_{k_m}}}{2} \sin \mleft( \varphi_{j_n, k_m} \mright) = \frac{1}{4i} \sum_{n = 1}^{M_j} \sum_{m = 1}^{M_k} \sqrt{\gamma_{j_n} \gamma_{k_m}} \mleft[ \exp \mleft( i \varphi_{j_n, k_m} \mright) - \exp \mleft( - i \varphi_{j_n, k_m} \mright) \mright] \nn\\
&=& \frac{1}{4i} \sum_{n = 1}^{M_j} \sum_{\substack{m = 1 \\ x_{j_{M_j}} < x_{k_m}}}^{M_k} \mleft[ \sqrt{\gamma_{j_n}} \exp \mleft( i \varphi_{j_n, j_{M_j}} \mright) \sqrt{\gamma_{k_m}} \exp \mleft( i \varphi_{j_{M_j}, k_m} \mright) - \text{H.c.} \mright] \nn\\
&& + \frac{1}{4i} \sum_{n = 1}^{M_j} \sum_{\substack{m = 1 \\ x_{j_{M_j}} > x_{k_m} > x_{j_n}}}^{M_k} \mleft[ \sqrt{\gamma_{j_n}} \exp \mleft( i \varphi_{j_n, j_{M_j}} \mright) \sqrt{\gamma_{k_m}} \exp \mleft( - i \varphi_{j_{M_j}, k_m} \mright) - \text{H.c.} \mright] \nn\\
&& + \frac{1}{4i} \sum_{n = 1}^{M_j} \sum_{\substack{m = 1 \\ x_{j_n} > x_{k_m}}}^{M_k} \mleft[ \sqrt{\gamma_{j_n}} \exp \mleft( - i \varphi_{j_n, j_{M_j}} \mright) \sqrt{\gamma_{k_m}} \exp \mleft( i \varphi_{j_{M_j}, k_m} \mright) - \text{H.c.} \mright] \nn\\
&=& \frac{1}{4i} \sum_{n = 1}^{M_j} \sum_{\substack{m = 1 \\ x_{j_{M_j}} < x_{k_m}}}^{M_k} \mleft[ \sqrt{\gamma_{j_n}} \exp \mleft( i \varphi_{j_n, j_{M_j}} \mright) \sqrt{\gamma_{k_m}} \exp \mleft( i \varphi_{j_{M_j}, k_m} \mright) - \text{H.c.} \mright] \nn\\
&& + \frac{1}{4i} \sum_{n = 1}^{M_j} \sum_{\substack{m = 1 \\ x_{j_{M_j}} > x_{k_m}}}^{M_k} \mleft[ \sqrt{\gamma_{j_n}} \exp \mleft( i \varphi_{j_n, j_{M_j}} \mright) \sqrt{\gamma_{k_m}} \exp \mleft( - i \varphi_{j_{M_j}, k_m} \mright) - \text{H.c.} \mright] \nn\\
&& + 2 \times \frac{1}{4i} \sum_{n = 1}^{M_j} \sum_{\substack{m = 1 \\ x_{j_n} > x_{k_m}}}^{M_k} \mleft[ \sqrt{\gamma_{j_n}} \exp \mleft( - i \varphi_{j_n, j_{M_j}} \mright) \sqrt{\gamma_{k_m}} \exp \mleft( i \varphi_{j_{M_j}, k_m} \mright) - \text{H.c.} \mright] \nn\\
&=& \frac{1}{4i} \underbrace{\sum_{n = 1}^{M_j} \sqrt{\gamma_{j_n}} \exp \mleft( i \varphi_{j_n, j_{M_j}} \mright)}_{= 0 \:\text{if}\: \Gamma_j = 0} \mleft[ \sum_{\substack{m = 1 \\ x_{j_{M_j}} < x_{k_m}}}^{M_k} \sqrt{\gamma_{k_m}} \exp \mleft( i \varphi_{j_{M_j}, k_m} \mright) - \text{H.c.} + \sum_{\substack{m = 1 \\ x_{j_{M_j}} > x_{k_m}}}^{M_k} \sqrt{\gamma_{k_m}} \exp \mleft( - i \varphi_{j_{M_j}, k_m} \mright) - \text{H.c.} \mright] \nn\\
&& + \sum_{n = 1}^{M_j} \sum_{\substack{m = 1 \\ x_{j_n} > x_{k_m}}}^{M_k} \sqrt{\gamma_{j_n} \gamma_{k_m}} \sin \mleft( \varphi_{j_n, k_m} \mright).
\label{eq:gjkGeneralConnection}
\eea

If the atoms are separate, $x_{j_{M_j}} < x_{k_1}$, and thus the second line of the final expression in \eqref{eq:gjkGeneralConnection} contain no terms, which means that $\Gamma_j = 0$ implies $g_{j, k} = 0$ in this case. 

If the atoms are nested, $x_{j_p} < x_{k_1} <  x_{k_{M_k}} < x_{j_{p+1}}$ ($p < M_j$), the second line of the final expression in \eqref{eq:gjkGeneralConnection} can be rewritten as
\bea
\sum_{n = 1}^{M_j} \sum_{\substack{m = 1 \\ x_{j_n} > x_{k_m}}}^{M_k} \sqrt{\gamma_{j_n} \gamma_{k_m}} \sin \mleft( \varphi_{j_n, k_m} \mright) = \sum_{n = p + 1}^{M_j} \sum_{m = 1}^{M_k} \sqrt{\gamma_{j_n} \gamma_{k_m}} \sin \mleft( \varphi_{j_n, k_m} \mright),
\eea
which we recognize as the expression for the interaction between two separate giant atoms, where the first giant atom is atom $k$ and the second giant atom is atom $j$ only connected at the connection points from $j_{p+1}$ to $j_{M_j}$. We can thus conclude that $\Gamma_j = 0$ and $\Gamma_k = 0$ together imply that $g_{j, k} = 0$ for the case of nested atoms.

If the atoms are braided, the second line of the final expression in \eqref{eq:gjkGeneralConnection} may be nonzero. The implications of setting all individual decay terms to zero in the three different geometries are thus the same for giant atoms with an arbitrary number of connection points as for giant atoms with two connection points.

As for the implications of setting the collective decay or the exchange interaction to zero, we have already shown for the case of giant atoms with two connection points that such a condition does not necessarily have to make any other terms zero. Thus, that is also true for giant atoms with an arbitrary number of connection points.

\bibliography{GiantAtomRefs_SuppMat}